\documentstyle[epsfig,12pt]{article}

\begin{document}
\begin{center}
{\Large\bf Covariant tensor formalism for partial wave analyses of
$\psi$ decay to mesons}
\vskip 0.5cm

{\large\bf B.S. Zou$^{a,b,}$\footnote{zoubs@mail.ihep.ac.cn} and
D.V.Bugg$^c$}

a) CCAST (World Laboratory), P.O.~Box 8730, Beijing 100080\\
b) Institute of High Energy Physics, CAS, P.O.~Box 918(4), Beijing 100039,
China\\
c) Queen Mary College, London, UK\\
\vskip 0.5cm
\end{center}

\begin{abstract}
$J/\psi$ and $\psi'$ decay to mesons are a good place to look for
glueballs, hybrids and for extracting strange and nonstrange
components in mesons. Abundant $J/\psi$ and $\psi'$ events have
been collected at the Beijing Electron Positron Collider (BEPC).
More data will be collected at upgraded BEPC and CLEO-C. Here we
provide explicit PWA formulae for many interesting channels in the
covariant tensor formalism.
\end{abstract}

\section{Introduction}

High statistics data have appeared from BES for $J/\psi$ decays
and will soon be available also for $\psi '$ decays. Further high
statistics data are expected from CLEO\cite{CLEO}. It is
convenient to have a uniform approach to partial wave analyses.
Here we provide one such approach using covariant tensor
formalism. A similar approach has been used in analyzing other
reactions\cite{Anisovich1,Anisovich2,Anisovich3}. We provide
formulae documenting those which have been used for a number of
channels already published by BES\cite{BES1,BES2,BES3,BES4,BES5}
and extend them to further channels being prepared for
publication. This list of reactions is not exhaustive, but
formulae are readily extended to other cases following the same
methods.

Reactions fall into two categories: non-radiative decays, where
final-state particles are pions or kaons; all polarization
information is then available in the form of angular
distributions. Reactions of this type are discussed in Section 2.
This formalism extends also to final states containing the
$\omega$, where polarization information is measured fully by the
decay $\omega \to \pi ^+ \pi ^- \pi ^0$. The second class of
reactions consists of radiative decays, e.g. $J/\psi  \to \gamma
\pi ^+\pi ^-$. For this class,  differential cross sections need
to be summed over the unmeasured helicities of the photon,
incorporating the knowledge that the photon is transverse. These
reactions are considered in Section 3.

\section{Formalism for $\psi$ non-radiative decay to mesons}
The general form for the decay amplitude of a vector meson $\psi$
with spin projection of $m$ is
\begin{equation}
A=\psi_\mu(m)A^\mu=\psi_\mu(m)\sum_i\Lambda_i U_i^{\mu}
\end{equation}
where $\psi_\mu(m)$ is the polarization vector of the $\psi$;
$U_i^{\mu}$ is the $i$-th partial wave amplitude with coupling
strength determined by a complex parameter $\Lambda_i$. The
polarization vector satisfies
\begin{equation}
\sum_{m=1}^3\psi^\mu(m)\psi^{*\nu}(m)= -g^{\mu\nu}+{p_\psi^\mu
p_\psi^\nu\over p_\psi^2}\equiv -\tilde
g^{\mu\nu}(p_\psi).
\end{equation}

For $\psi$ production from $e^+ e^-$ annihilation, the electrons
are highly relativistic, with the result that $J_z = \pm 1$. If we
take the beam direction to the be the z-axis, this limits $m$ to 1
and 2, i.e. components along $x$ and $y$. Then the differential
cross section for decay to an $n$-body final state is:

\begin{equation}
\frac{d\sigma}{d\Phi_n} =
\frac{(2\pi)^4}{2M_\psi}\cdot\frac{1}{2}\sum^2_{m=1} \psi_{\mu}(m)
A^{\mu}\psi^*_{\mu'}(m) A^{*\mu'}
\end{equation}
where $M_\psi$ is the mass of $\psi$ and $d\Phi_n$ is the standard
element of n-body phase space given by
\begin{equation}
d\Phi_n(p_\psi; p_1, \cdots p_n)=\delta^4(p_\psi-\sum^n_{i=1}p_i)
\prod^n_{i=1}\frac{d^3{\bf p}_i}{(2\pi)^32E_i}.
\end{equation}

Note that
\begin{equation}
\sum^2_{m=1}\psi_{\mu}(m)\psi^*_{\mu'}(m)
=\delta_{\mu\mu'}(\delta_{\mu 1}+\delta_{\mu 2}),
\end{equation}
so we have
\begin{equation}
\frac{d\sigma}{d\Phi_n} = \frac{1}{2}\sum^2_{\mu=1}A^{\mu}A^{*\mu}
= \frac{1}{2}\sum_{i,j}\Lambda_i\Lambda_j^*\sum^2_{\mu=1}
U_i^{\mu}U_j^{*\mu}
 \equiv \sum_{i,j}P_{ij}\cdot F_{ij}
\end{equation}
where
\begin{eqnarray}
P_{ij} &= P^*_{ji} &= \Lambda_i\Lambda^*_j, \\
F_{ij} &= F^*_{ji} &= \frac{1}{2}\sum^2_{\mu=1}
U_i^{\mu}U_j^{*\mu}.
\end{eqnarray}

We construct the partial wave amplitudes $U_i^{\mu}$ in the
covariant Rarita-Schwinger tensor formalism \cite{Rarita}. As in
Ref. \cite{Chung}, we use pure orbital angular momentum covariant
tensors $\tilde t^{(l)}_{\mu_1\cdots\mu_l}$  and covariant spin
wave functions $\phi_{\mu_1\cdots\mu_S}$ together with operators
$g_{\mu\nu}$, $\epsilon_{\mu\nu\lambda\sigma}$ and momenta of
parent particles. For a process $a\to bc$, the covariant tensors
$\tilde t^{(l)}_{\mu_1\cdots\mu_l}$ for final states of pure
orbital angular momentum $l$ are constructed from relevant momenta
$p_a$, $p_b$ and $p_c$ \cite{Chung}
\begin{eqnarray}
\tilde t^{(0)} &=& 1 , \\
\tilde t^{(1)}_\mu &=& \tilde g_{\mu\nu}(p_a)r^\nu B_1(Q_{abc})
\equiv\tilde r_\mu B_1(Q_{abc}),\\
\tilde t^{(2)}_{\mu\nu} &=& [\tilde r_\mu\tilde r_\nu
-{1\over 3}(\tilde r\cdot\tilde r)\tilde g_{\mu\nu}(p_a)]B_2(Q_{abc}), \\
\tilde t^{(3)}_{\mu\nu\lambda} &=& [\tilde r_\mu\tilde r_\nu\tilde
r_\lambda -{1\over 5}(\tilde r\cdot\tilde r)(\tilde
g_{\mu\nu}(p_a)\tilde r_\lambda +\tilde g_{\nu\lambda}(p_a)\tilde
r_\mu+\tilde g_{\lambda\mu}(p_a)\tilde r_\nu)]B_3(Q_{abc}),
\\
 & & \cdots  \nonumber
\end{eqnarray}
with $r=p_b-p_c$. The term $(\tilde r\cdot\tilde r)$ is the
dot-product of 4-vectors: $\tilde r_0 \tilde r_0 - \tilde r_1
\tilde r_1 - \tilde r_2 \tilde r_2 - \tilde r_3 \tilde r_3$, and
makes $\tilde t^{(2)}_{\mu \nu }$ traceless. Likewise $\tilde
t^{(3)}$ is constructed to be traceless. $Q_{abc}$ is the
magnitude of $\bf p_b$ or $\bf p_c$ in the rest system of $a$,
where
\begin{equation}
Q_{abc}^2=\frac{(s_{a}+s_{b}-s_{c})^{2}}{4s_{a}}-s_{b}
\end{equation}
with $s_a=E_a^2-{\bf p}^2_a$. Then $\tilde
t^{(l)}_{\mu_1\cdots\mu_l}$ contains the angular distribution
function multiplied by a Blatt-Weisskopf barrier
factor\cite{Chung,Hippel} $Q_{abc}^lB_l(Q_{abc})$. Explicitly
\begin{eqnarray}
B_1(Q_{abc})\!&=&\! \sqrt{2\over Q_{abc}^2+Q^2_0} ,\\
B_2(Q_{abc})\!&=&\! \sqrt{13\over Q_{abc}^4+3Q_{abc}^2Q^2_0+9Q_0^4}, \\
B_3(Q_{abc})\!&=&\! \sqrt{277\over
Q_{abc}^6+6Q_{abc}^4Q^2_0+45Q_{abc}^2Q_0^4
                    +225Q_0^6} , \\
B_4(Q_{abc})\!&=&\! \sqrt{12746\over
Q_{abc}^8+10Q_{abc}^6Q_0^2+135Q_{abc}^4Q^4_ 0
                    +1575Q_{abc}^2Q_0^6+11025Q_0^8}
\end{eqnarray}
Here $Q_0$ is a hadron ``scale" parameter $Q_0 = 0.197321/R$
GeV/c, where $R$ is the radius of the centrifugal barrier in fm.
We remark that in these Blatt-Weisskopf factors, the approximation
is made that the centrifugal barrier may be replaced by a square
well of radius $R$.

If $a$ is an intermediate resonance decaying into $bc$, one needs
to introduce into the amplitude a Breit-Wigner propagator denoted
by $f^{(a)}_{(bc)}$:
\begin{equation}
f^{(a)}_{(bc)}={1\over m_a^2-s_{bc}-im_a\Gamma_a};
\end{equation}
here $s_{bc}=(p_b+p_c)^2$ is the invariant mass-squared of $b$ and
$c$; $m_a$, $\Gamma_a$ are the resonance mass and width.

We outline now some further general features of notation, taking
as an example the two-step process $J/\psi \to \rho _{12}\pi _3$,
$\rho _{12} \to \pi _1\pi _2$. In the first step we denote the
orbital angular momentum by $L$; in this example $L= 1$. In the
second step, we denote the orbital angular momentum by $\ell $,
which is again 1 in this case. The tensor describing the first
step will be denoted by $\tilde T^{(L)}_{\mu_1\cdots\mu_L}$. The
tensor describing the second step will be denoted by $\tilde
t^{(l)}_{\mu_1\cdots\mu_l}$. The orbital angular momentum is
constructed in terms of relative momenta, so it is convenient to
define $q_{(ij)}=p_i-p_j$.

Some expressions depend also on the total momentum of the $ij$
pair: $p_{(ij)}=p_i+p_j$. When one wants to combine two angular
momenta (${\bf j}_b$ and ${\bf j}_c$) into a total angular
momentum ${\bf j}_a$, if $j_a+j_b+j_c$ is an odd number, then a
combination $\epsilon_{\mu\nu\lambda\sigma}p_a^\mu$ with $p_a$ the
momentum of the parent particle is needed; otherwise it is not
needed.

Projection operators will be a useful general tool in constructing
expressions. For a meson $a$ with spin $S$ and corresponding spin
wave function $\phi_{\mu_1\cdots\mu_S}(p_a,m)$, what we usually
need to use in constructing amplitudes is its spin projection
operator $P^{(S)}_{\mu_1\cdots\mu_S\mu'_1\cdots\mu'_S}(p_a)$.
\begin{eqnarray}
P^{(1)}_{\mu\mu'}(p_a) \!&=&\!
\sum_m\phi_\mu(p_a,m)\phi^*_{\mu'}(p_a,m)=-g_{\mu\mu'}+{p_{a\mu}p_{a\mu'}\over
p_a^2}\equiv -\tilde g_{\mu\mu'}(p_a), \\
P^{(2)}_{\mu\nu\mu'\nu'}(p_a) \!&=&\!
\sum_m\phi_{\mu\nu}(p_a,m)\phi^*_{\mu'\nu'}(p_a,m)={1\over 2}
(\tilde g_{\mu\mu'}\tilde g_{\nu\nu'}+\tilde g_{\mu\nu'}\tilde
g_{\nu\mu'})-{1\over 3}\tilde g_{\mu\nu}\tilde g_{\mu'\nu'}, \\
P^{(3)}_{\mu\nu\lambda\mu'\nu'\lambda'}(p_a) \!&=&\!
\sum_m\phi_{\mu\nu\lambda}(p_a,m)\phi^*_{\mu'\nu'\lambda'}(p_a,m)\nonumber\\
&=&\! -{1\over 6} (\tilde g_{\mu\mu'}\tilde g_{\nu\nu'}\tilde
g_{\lambda\lambda'}+\tilde g_{\mu\mu'}\tilde g_{\nu\lambda'}\tilde
g_{\lambda\nu'}+\tilde g_{\mu\nu'}\tilde g_{\nu\mu'}\tilde
g_{\lambda\lambda'} \nonumber\\
& & \quad\quad +\tilde g_{\mu\nu'}\tilde g_{\nu\lambda'}\tilde
g_{\lambda\mu'}+\tilde g_{\mu\lambda'}\tilde g_{\nu\nu'}\tilde
g_{\lambda\mu'}+\tilde g_{\mu\lambda'}\tilde g_{\nu\mu'}\tilde
g_{\lambda\nu'})\nonumber\\
& &\! +{1\over 15}(\tilde g_{\mu\nu}\tilde g_{\mu'\nu'}\tilde
g_{\lambda\lambda'}+\tilde g_{\mu\nu}\tilde g_{\nu'\lambda'}\tilde
g_{\lambda\mu'}+\tilde g_{\mu\nu}\tilde g_{\mu'\lambda'}\tilde
g_{\lambda\nu'}\nonumber\\
& & \quad\quad +\tilde g_{\mu\lambda}\tilde g_{\mu'\lambda'}\tilde
g_{\nu\nu'}+\tilde g_{\mu\lambda}\tilde g_{\mu'\nu'}\tilde
g_{\nu\lambda'}+\tilde g_{\mu\lambda}\tilde g_{\nu'\lambda'}\tilde
g_{\nu\mu'}\nonumber\\
& & \quad\quad +\tilde g_{\nu\lambda}\tilde g_{\nu'\lambda'}\tilde
g_{\mu\mu'}+\tilde g_{\nu\lambda}\tilde g_{\mu'\nu'}\tilde
g_{\mu\lambda'}+\tilde g_{\nu\lambda}\tilde g_{\mu'\lambda'}\tilde
g_{\mu\nu'}), \\
P^{(4)}_{\mu\nu\lambda\sigma\mu'\nu'\lambda'\sigma'}(p_a) \!&=&\!
\sum_m\phi_{\mu\nu\lambda\sigma}(p_a,m)\phi^*_{\mu'\nu'\lambda'\sigma'}(p_a,m)
\nonumber\\
&=&\! {1\over 24} [\tilde g_{\mu\mu'}\tilde g_{\nu\nu'}\tilde
g_{\lambda\lambda'}\tilde g_{\sigma\sigma'} +\cdots
 (\mu', \nu', \lambda', \sigma' ~permutation, ~24~terms)] \nonumber\\
& & -{1\over 84}[\tilde{g}_{\mu\nu}\tilde{g}_{\mu'\nu'}
\tilde{g}_{\lambda\lambda'}\tilde{g}_{\sigma\sigma'}+\cdots
 (\mu, \nu, \lambda, \sigma~permutation, \nonumber\\
& & \quad\quad\quad\quad\quad ~\mu', \nu', \lambda', \sigma'
~permutation, ~72~terms)] \nonumber\\
& & +{1\over 105}(\tilde{g}_{\mu\nu}\tilde{g}_{\lambda\sigma} +
\tilde{g}_{\mu\lambda}\tilde{g}_{\nu\sigma} +
\tilde{g}_{\mu\sigma}\tilde{g}_{\nu\lambda})
(\tilde{g}_{\mu'\nu'}\tilde{g}_{\lambda'\sigma'} +
\tilde{g}_{\mu'\lambda'}\tilde{g}_{\nu'\sigma'} +
\tilde{g}_{\mu'\sigma'}\tilde{g}_{\nu'\lambda'}). \nonumber\\
\end{eqnarray}

Note that
\begin{equation}
\tilde t^{(L)}_{\mu_1\cdots\mu_L}
=(-1)^LP^{(L)}_{\mu_1\cdots\mu_L\mu'_1\cdots\mu'_L}r^{\mu'_1}\cdots
r^{\mu'_L}B_L(Q_{abc}).
\end{equation}

We come now to specific examples of reactions.

\subsection{$\psi\to\pi^+\pi^-\pi^0$}
For three isospin 1 particles coupling to an isospin zero
particle, the only possible coupling for isospin conservation is
$\bf (I_1\times I_2)\cdot I_3$, which is fully anti-symmetric in
particles  1,2,3. This demands that the angular dependent part
should also be fully anti-symmetric for 1,2,3, in order to make
the overall amplitude symmetric. For $\psi\to\pi^+\pi^-\pi^0$, any
two pions are limited to an overall isospin 1 and hence can only
be negative parity states with $J$ odd, {\sl i.e.}, $J^P=1^-$,
$3^-$, $5^-$ {\sl etc}.

For $\psi\to\rho(1^-)\pi\to\pi^+\pi^-\pi^0$, $\psi$ decays to
$\rho\pi$ in a P-wave; then $\rho$ decays to $\pi\pi$ also in
P-wave, hence the amplitude for the two step process is
\begin{eqnarray}
U_\rho^\mu \!&=&\! {\bf (I_1\times I_2)\cdot I_3}
~\epsilon_{\mu\nu\lambda\sigma}p_\psi^\sigma \tilde
T^{(1)\nu}_{(\rho 3)}\tilde t^{(1)\lambda}_{(12)}
f^{(\rho)}_{(12)} + (1\leftrightarrow 3) +
(2\leftrightarrow 3) \nonumber\\
&=&\! 4i\epsilon_{\mu\nu\lambda\sigma}p_1^\nu p_2^\lambda
p_3^\sigma \Bigl[B_1(Q_{\psi\rho 3}) f^{(\rho)}_{(12)}B_1(Q_{\rho
12}) +
B_1(Q_{\psi\rho 2}) f^{(\rho)}_{(13)} B_1(Q_{\rho 13}) \nonumber\\
& &\quad\quad\quad + B_1(Q_{\psi\rho 1}) f^{(\rho)}_{(23)}
B_1(Q_{\rho 23})\Bigr].
\end{eqnarray}
Here we use the convention ${\bf I}_1=({-1\over\sqrt{2}},
{-i\over\sqrt{2}},0)$ for $\pi^+$, ${\bf I}_2=({1\over\sqrt{2}},
{-i\over\sqrt{2}},0)$ for $\pi^-$ and ${\bf I}_3 = (0, 0, 1)$ for
$\pi^0$. This gives ${\bf (I_1\times I_2)\cdot I_3}=-i$.

The amplitude can be further simplified in the $\psi$ rest system
as
\begin{eqnarray}
U_\rho^\mu &=& 4iM_\psi\epsilon_{\mu\nu\lambda 0}p_1^\nu
p_2^\lambda \Bigl[B_1(Q_{\psi\rho 3}) f^{(\rho)}_{(12)}B_1(Q_{\rho
12}) +
B_1(Q_{\psi\rho 2}) f^{(\rho)}_{(13)} B_1(Q_{\rho 13}) \nonumber\\
& &\quad\quad\quad + B_1(Q_{\psi\rho 1}) f^{(\rho)}_{(23)}
B_1(Q_{\rho 23})\Bigr].
\end{eqnarray}

For any other $1^-$ intermediate state $\rho'$, one can get the
corresponding amplitude by simply replacing the Breit-Wigner
component $f^{(\rho)}$  by $f^{(\rho')}$.

For $\psi\to\rho_3(3^-)\pi\to\pi^+\pi^-\pi^0$, $\psi$ decays to
$\rho_3\pi$ in F-wave; then $\rho_3$ decays to $\pi\pi$ also in
F-wave; the amplitude is
\begin{eqnarray} U_{\rho_3}^\mu \!&=&\! {\bf (I_1\times I_2)\cdot
I_3} ~\epsilon_{\mu\nu\lambda\sigma}p_\psi^\sigma \tilde
T^{(3)\nu\alpha\beta}_{(\rho_3 3)}\tilde
t^{(3)\lambda}_{(12)\alpha\beta}\cdot f^{(\rho_3)}_{(12)} +
(1\leftrightarrow 3) +
(2\leftrightarrow 3) \nonumber\\
&=&\! -i\epsilon_{\mu\nu\lambda\sigma}p_\psi^\sigma [\tilde
T^{(3)\nu\alpha\beta}_{(\rho_3 3)}\tilde
t^{(3)\lambda}_{(12)\alpha\beta}\cdot f^{(\rho_3)}_{(12)} -
(1\leftrightarrow 3) - (2\leftrightarrow 3)].
\end{eqnarray}

Similarly, for $\psi\to\rho_5(5^-)\pi\to\pi^+\pi^-\pi^0$, the
amplitude should be
\begin{eqnarray}
U_{\rho_5}^\mu \!&=&\! {\bf (I_1\times I_2)\cdot I_3}
~\epsilon_{\mu\nu\lambda\sigma}p_\psi^\sigma \tilde
T^{(5)\nu\alpha\beta\gamma\delta}_{(\rho_53)} \tilde
t^{(5)\lambda}_{(12)\alpha\beta\gamma\delta}f^{(\rho_5)}_{(12)} +
(1\leftrightarrow 3) +
(2\leftrightarrow 3) \nonumber\\
&=&\! -i\epsilon_{\mu\nu\lambda\sigma}p_\psi^\sigma [\tilde
T^{(5)\nu\alpha\beta\gamma\delta}_{(\rho_53)} \tilde
t^{(5)\lambda}_{(12)\alpha\beta\gamma\delta}f^{(\rho_5)}_{(12)} -
(1\leftrightarrow 3) - (2\leftrightarrow 3)].
\end{eqnarray}

If one considers a small isospin symmetry breaking effect, a free
parameter can be multiplied into the term corresponding to the
$\rho^0$ intermediate state.

\subsection{$\psi\to K^+K^-\pi^0$}
This channel is similar to $\pi^+\pi^-\pi^0$. However, we now need
to consider resonances for both $K\pi$ and $K^+K^-$ subsystems.
Numbering $K^+$, $K^-$, $\pi^0$ as particle 1, 2, 3, the possible
partial wave amplitudes are the following:
\begin{eqnarray}
U_{\rho'}^\mu \!&=&\! \epsilon_{\mu\nu\lambda\sigma} p_1^\nu
p_2^\lambda p_3^\sigma B_1(
Q_{\psi\rho' 3}) f^{(\rho')}_{(12)}B_1(Q_{\rho' 12}), \\
U_{\rho_3}^\mu \!&=&\! \epsilon_{\mu\nu\lambda\sigma}p_\psi^\sigma
\tilde T^{(3)\nu\alpha\beta}_{(\rho_3 3)}\tilde
t^{(3)\lambda}_{(12)\alpha\beta}\cdot f^{(\rho_3)}_{(12)},\\
U_{\rho_5}^\mu \!&=&\! \epsilon_{\mu\nu\lambda\sigma}
p_\psi^\sigma \tilde T^{(5)\nu\alpha\beta\gamma\delta}_{(\rho_53)}
\tilde t^{(5)\lambda}_{(12)\alpha\beta\gamma\delta}
f^{(\rho_5)}_{(12)},\\
U_{K^*}^\mu \!&=&\! \epsilon_{\mu\nu\lambda\sigma} p_1^\nu
p_2^\lambda p_3^\sigma \Bigl[B_1(Q_{\psi K^* 2}) f^{(K^*)}_{(13)}
B_1(Q_{K^* 13}) +
B_1(Q_{\psi K^* 1}) f^{(K^*)}_{(23)} B_1(Q_{K^* 23})\Bigr], \\
U_{K^*_3}^\mu \!&=&\! \epsilon_{\mu\nu\lambda\sigma} p_\psi^\sigma
[\tilde T^{(3)\nu\alpha\beta}_{(K^*_3 2)}\tilde
t^{(3)\lambda}_{(13)\alpha\beta}\cdot f^{(K^*_3)}_{(13)} -
(1\leftrightarrow 2)], \\
U_{K^*_5}^\mu \!&=&\! \epsilon_{\mu\nu\lambda\sigma} p_\psi^\sigma
[\tilde T^{(5)\nu\alpha\beta\gamma\delta}_{(K^*_52)} \tilde
t^{(5)\lambda}_{(13)\alpha\beta\gamma\delta}f^{(K^*_5)}_{(13)} -
(1\leftrightarrow 2)].
\end{eqnarray}

\subsection{$\psi\to\phi\pi^+\pi^-\to K^+K^-\pi^+\pi^-$}

For this channel, $\phi$ is reconstructed from two kaons; most
possible intermediate states are $\phi$ plus an isospin zero
resonance, $f_0$ or $f_2$, which decays into two pions. The $f_4$
is unlikely to be produced, because $\psi$ mass is not far from
$\phi f_4$ threshold and the decay requires $L=2$ between $\phi$
and $f_4$, hence a strong centrifugal barrier. For $\psi\to\phi
f_J$ in an orbital angular momentum $L$ state, the conservation of
total angular momentum requires
\begin{equation}
\bf S_\psi = S+L
\end{equation}
where
\begin{equation}
\bf S=S_\phi + J.
\end{equation}
In the following, we use notation $<\phi f_J|LS>$ to denote the
corresponding partial wave amplitude $U_i^\mu$. We number the
$K^+$, $K^-$, $\pi^+$, $\pi^-$ as particle 1, 2, 3, 4,
respectively. Then we have two independent partial wave amplitudes
for each $f_0$ production. In the general formalism, they may be
written:
\begin{eqnarray}
<\phi f_0|01> \!&=&\! \tilde t^{(1)\mu}_{(12)}f_{(12)}^{(\phi)}f_{(34)}^{(f_0)}, \\
<\phi f_0|21> \!&=&\! \tilde T^{(2)\mu\nu}_{(\phi f_0)}\tilde
t^{(1)}_{(12)\nu}f_{(12)}^{(\phi)} f_{(34)}^{(f_0)}.
\end{eqnarray}
For the very narrow $\phi$ resonance, the $\ell = 1$ centrifugal
barrier factor for $\phi$ decay has negligible effect on the
$\phi$ line-shape and can be dropped. The expression for
$t^{(1)\mu}_{(12)}$ simplifies to
$$t^{(1)\mu}_{(12)} = \tilde r^\mu=q_{(12)}^\mu.$$
In the last step, we use the fact that $K^+$ and $K^-$ have equal
masses. Then Eqs. (36) and (37) become:
\begin{eqnarray}
<\phi f_0|01> \!&=&\! q^\mu_{(12)} f_{(12)}^{(\phi)}f_{(34)}^{(f_0)}, \\
<\phi f_0|21> \!&=&\! \tau ^{\mu \nu}q_{(12)\nu}
            B_2(Q_{\psi\phi f_0})f_{(12)}^{(\phi)} f_{(34)}^{(f_0)},
\end{eqnarray}
where $\tau ^{\mu \nu}$ is the $L = 2$ operator
\begin {equation}
\tau ^{\mu \nu } = q^\mu_{(12)} q^\nu_{(12)} - \frac
{1}{3}(q_{(12)}\cdot q_{(12)}) g^{\mu \nu }.
\end {equation}

For each $f_2$ production, there are five independent partial
waves, which we retain in their general form:
\begin{eqnarray}
<\phi f_2|01> \!&=&\! \tilde t^{(2)\mu\nu}_{(34)} \tilde
t^{(1)}_{(12)\nu}f_{(12)}^{(\phi)}f_{(34)}^{(f_2)} , \\
<\phi f_2|21> \!&=&\! \tilde T^{(2)\mu\alpha}_{(\phi f_2)} \tilde
t^{(2)}_{(34)\alpha\nu}\tilde t^{(1)\nu}_{(12)}
        f_{(12)}^{(\phi)} f_{(34)}^{(f_2)}, \\
<\phi f_2|22> \!&=&\! \epsilon^{\mu\alpha\beta\gamma}
p_{\psi\alpha} \tilde T^{(2)\delta}_{(\phi f_2)\beta}
[\epsilon_{\gamma\lambda\sigma\nu} \tilde
t^{(2)\lambda}_{(34)\delta} +\epsilon_{\delta\lambda\sigma\nu}
\tilde t^{(2)\lambda}_{(34)\gamma}]p_\psi^\sigma \tilde
t^{(1)\nu}_{(12)}f_{(12)}^{(\phi)}f_{(34)}^{(f_2)}, \\
<\phi f_2|23> \!&=&\! P^{(3)\mu\alpha\beta\gamma\delta\nu}(p_\psi)
\tilde T^{(2)}_{(\phi f_2)\alpha\beta} \tilde
t^{(2)}_{(34)\gamma\delta} \tilde t^{(1)}_{(12)\nu}
f_{(12)}^{(\phi)}f_{(34)}^{(f_2)}, \\
<\phi f_2|43> \!&=&\! \tilde T^{(4)\mu\nu\lambda\sigma}_{(\phi
f_2)}
    \tilde t^{(1)}_{(12)\nu} \tilde t^{(2)}_{(34)\lambda\sigma}
        f_{(12)}^{(\phi)}f_{(34)}^{(f_2)}.
\end{eqnarray}

There is no established resonance decaying into $\phi\pi$.
However, there are speculations about ($s\bar sq\bar q$)
four-quark states which could decay to $\phi\pi$. So here we also
give some partial wave amplitudes for $\psi\to X\pi$ with the
intermediate resonance $X$ further decaying to $\phi\pi$. For $X$
being a  $\rho'(1^{--})$ state, there is only one independent
amplitude since both $\psi\to\rho'\pi$ and $\rho'\to\phi\pi$ are
limited to a $P$ wave.
\begin{eqnarray}
U^\mu_{\rho'} \!&=&\!
\epsilon^\mu_{\alpha\beta\gamma}p_\psi^\alpha
        [\tilde T^{(1)\beta}_{(\rho'3)}
\epsilon^{\gamma\delta\sigma\lambda}p_{\psi\delta} \tilde
t^{(1)}_{(\phi 4)\sigma} \tilde t^{(1)}_{(12)\lambda}
            f_{(12)}^{(\phi)} f_{(\phi 4)}^{(\rho')}
 + \tilde T^{(1)\beta}_{(\rho'4)}
\epsilon^{\gamma\delta\sigma\lambda}p_{\psi\delta}\tilde
t^{(1)}_{(\phi 3)\sigma} \tilde t^{(1)}_{(12)\lambda}
            f_{(12)}^{(\phi)} f_{(\phi 3)}^{(\rho')}].
\end{eqnarray}
For $X$ being a $b_1(1^{+-})$ state, there are four independent amplitudes
since both $\psi\to b_1\pi$ and $b_1\to\phi\pi$ can have both S and D waves.
\begin{eqnarray}
U^\mu_{b_1SS}  \!&=&\! \tilde g_{(123)}^{\mu\nu}\tilde
t^{(1)}_{(12)\nu}
            f_{(12)}^{(\phi)} f_{(123)}^{(b_1)} +
                 \tilde g_{(124)}^{\mu\nu}\tilde t^{(1)}_{(12)\nu}
            f_{(12)}^{(\phi)} f_{(124)}^{(b_1)} ,\\
U^\mu_{b_1SD} \!&=&\! \tilde t_{(\phi3)}^{(2)\mu\nu}\tilde
t^{(1)}_{(12)\nu}
            f_{(12)}^{(\phi)} f_{(123)}^{(b_1)} +
                 \tilde t_{(\phi 4)}^{(2)\mu\nu}\tilde t^{(1)}_{(12)\nu}
            f_{(12)}^{(\phi)} f_{(124)}^{(b_1)} ,\\
U^\mu_{b_1DS} \!&=&\! \tilde T_{(b_14)}^{(2)\mu\lambda}\tilde
g_{(123)\lambda\nu}
            \tilde t^{(1)\nu}_{(12)}
            f_{(12)}^{(\phi)} f_{(123)}^{(b_1)} +
                 \tilde T_{(b_13)}^{(2)\mu\lambda} \tilde g_{(124)\lambda\nu}
              \tilde t^{(1)\nu}_{(12)}
            f_{(12)}^{(\phi)} f_{(124)}^{(b_1)} ,\\
U^\mu_{b_1DD} \!&=&\! \tilde T_{(b_14)}^{(2)\mu\lambda} \tilde
t^{(2)}_{(\phi 3)\lambda\nu}\tilde t^{(1)\nu}_{(12)}
f_{(12)}^{(\phi)} f_{(123)}^{(b_1)}+ \tilde
T_{(b_13)}^{(2)\mu\lambda} \tilde t^{(2)}_{(\phi 4)\lambda\nu}
\tilde t^{(1)\nu}_{(12)} f_{(12)}^{(\phi)} f_{(124)}^{(b_1)} .
\end{eqnarray}

\subsection{$\psi\to\omega K^+K^-\to\pi^+\pi^-\pi^0K^+K^-$}

The formulae for this channel are quite similar to those in the
previous subsection for the $\psi\to\phi\pi^+\pi^-$. If we number
the $\pi^0$, $\pi^+$, $\pi^-$, $K^+$, $K^-$ as 0, 1, 2, 3, 4, then
we can get corresponding partial wave amplitudes by simply
replacing $\tilde t^{(1)\mu}_{(12)}$ in equations of the previous
subsection by $\omega^\mu$ defined as
\begin{equation}
\omega^\mu \!=\! \epsilon^\mu_{\nu\lambda\sigma}p_1^\nu p_2^\lambda p_0^\sigma
\Bigl[B_1(Q_{\omega\rho 0}) f^{(\rho)}_{(12)}B_1(Q_{\rho 12}) +
B_1(Q_{\omega\rho 2}) f^{(\rho)}_{(10)} B_1(Q_{\rho 10}) +
B_1(Q_{\omega\rho 1}) f^{(\rho)}_{(20)} B_1(Q_{\rho 20})\Bigr]
\end{equation}
and replacing $f_{(12)}^{(\phi)}$ by $f_{(012)}^{(\omega)}$.

\subsection{$\psi\to K^+\pi^-K^-\pi^+$}

We label $K^+$,$\pi^-$,$K^-$, $\pi^+$ as 1,2,3,4. For $\rho a_0$
and $\rho a_2$ intermediate states, the formulae are the same as
for $\phi f_0$ and $\phi f_2$ intermediate states with a trivial
exchange between pions and kaons. For $KK^{*'}\to(KK^*\pi~or~K\rho
K)$, or $\pi\rho'\to K^*K$ intermediate states, the formulae are
the same as for the $\pi\rho'\to\pi\phi\pi$ intermediate state
with proper recombination of particles. For $KK^*_1\to
KK^*\pi~or~K\rho K)$ intermediate states, the formulae are the
same as for the $\pi b_1\to\pi\phi\pi$ intermediate state with
proper recombination of particles. So all the formulae given in
the subsection on $\phi\pi^+\pi^-$ may be applied here. In
addition, there are many more possible intermediate states. We
list additional formulae for some obvious large intermediate
states. Note for a resonance with the negative C-parity, it decays
to $K^*_{j_1}\bar K^*_{j_2}$ with a relative minus sign to its
charge conjugate  state $\bar K^*_{j_1}K^*_{j_2}$.
\begin{eqnarray}
<K^*K_0^*|01> \!&=&\! \tilde
t^{(1)\mu}_{(12)}f_{(12)}^{(K^*)}f_{(34)}^{(\bar K^*_0)}
-\tilde t^{(1)\mu}_{(34)}f_{(34)}^{(\bar K^*)}f_{(12)}^{(K^*_0)}, \\
<K^* K^*_0|21> \!&=&\! \tilde T^{(2)\mu\nu}_{((12)(34))} [\tilde
t^{(1)}_{(12)\nu}f_{(12)}^{(K^*)} f_{(34)}^{(K^*_0)} - \tilde
t^{(1)}_{(34)\nu}
f_{(34)}^{(K^*)} f_{(12)}^{(K^*_0)}],\\
<K^*K^*_2|01> \!&=&\! \tilde t^{(2)\mu\nu}_{(34)} \tilde
t^{(1)}_{(12)\nu}f_{(12)}^{(K^*)}f_{(34)}^{(K^*_2)} - \tilde
t^{(2)\mu\nu}_{(12)} \tilde t^{(1)}_{(34)\nu}
f_{(34)}^{(K^*)}f_{(12)}^{(K^*_2)}, \\
<K^*K^*_2|21> \!&=&\! \tilde T^{(2)\mu\alpha}_{((12)(34))} [\tilde
t^{(2)}_{(34)\alpha\nu}\tilde t^{(1)\nu}_{(12)}
f_{(12)}^{(K^*)}f_{(34)}^{(K^*_2)}  - \tilde
t^{(2)}_{(12)\alpha\nu}\tilde t^{(1)\nu}_{(34)}
f_{(34)}^{(K^*)}f_{(12)}^{(K^*_2)}],\\
<K^*K^*_2|22> \!&=&\! \epsilon^{\mu\alpha\beta\gamma}
p_{\psi\alpha} \tilde T^{(2)\delta}_{((12)(34))\beta}
\{[\epsilon_{\gamma\lambda\sigma\nu} \tilde
t^{(2)\lambda}_{(34)\delta} +\epsilon_{\delta\lambda\sigma\nu}
\tilde t^{(2)\lambda}_{(34)\gamma}]p_\psi^\sigma \tilde
t^{(1)\nu}_{(12)}
f_{(12)}^{(K^*)}f_{(34)}^{(K^*_2)} \nonumber\\
&& - [\epsilon_{\gamma\lambda\sigma\nu} \tilde
t^{(2)\lambda}_{(12)\delta} +\epsilon_{\delta\lambda\sigma\nu}
\tilde t^{(2)\lambda}_{(12)\gamma}]p_\psi^\sigma \tilde
t^{(1)\nu}_{(34)}f_{(34)}^{(K^*)}f_{(12)}^{(K^*_2)}\},\\
<K^*K^*_2|23> \!&=&\! P^{(3)\mu\alpha\beta\gamma\delta\nu}(p_\psi)
\tilde T^{(2)}_{((12)(34))\alpha\beta} [\tilde
t^{(2)}_{(34)\gamma\delta} \tilde t^{(1)}_{(12)\nu}
f_{(12)}^{(K^*)}f_{(34)}^{(K^*_2)} -\tilde
t^{(2)}_{(12)\gamma\delta} \tilde t^{(1)}_{(34)\nu}
f_{(34)}^{(K^*)}f_{(12)}^{(K^*_2)}], \nonumber \\ & & \\
<K^*K^*_2|43> \!&=&\! \tilde
T^{(4)\mu\nu\lambda\sigma}_{((12)(34))} [\tilde t^{(1)}_{(12)\nu}
\tilde t^{(2)}_{(34)\lambda\sigma}
f_{(12)}^{(K^*)}f_{(34)}^{(K^*_2)}-\tilde t^{(1)}_{(34)\nu} \tilde
t^{(2)}_{(12)\lambda\sigma}
f_{(34)}^{(K^*)}f_{(12)}^{(K^*_2)}],\\
<K^*_0K^{*'}_0|10> \!&=&\! \tilde T^{(1)\mu}_{((12)(34))}
[f_{(12)}^{K^*_0}f_{(34)}^{K^{*'}_0} +
f_{(34)}^{K^{*}_0}f_{(12)}^{K^{*'}_0}], \\
<K^*_0K^*_2|12> \!&=&\! \tilde T^{(1)}_{((12)(34))\nu}[\tilde
t^{(2)\mu\nu}_{(34)}f^{(K^*_0)}_{(12)}f^{(K^*_2)}_{(34)} +\tilde
t^{(2)\mu\nu}_{(12)}f^{(K^*_0)}_{(34)}f^{(K^*_2)}_{(12)}],\\
<K^*_0K^*_2|32> \!&=&\! \tilde T^{(3)\mu\nu\lambda}_{((12)(34))}
[\tilde t^{(2)}_{(34)\nu\lambda}
f^{(K^*_0)}_{(12)}f^{(K^*_2)}_{(34)} + \tilde
t^{(2)}_{(12)\nu\lambda}f^{(K^*_0)}_{(34)}
f^{(K^*_2)}_{(12)}],\\
<K^*K^{*'}|10> \!&=&\! \tilde T^{(1)\mu}_{((12)(34))} \tilde
t^{(1)\alpha}_{(12)} \tilde t^{(1)}_{(34)\alpha}
[f_{(12)}^{K^*}f_{(34)}^{K^{*'}} +
f_{(34)}^{K^*}f_{(12)}^{K^{*'}}],\\
<K^*K^{*'}|11> \!&=&\! \epsilon^{\mu\alpha\beta\gamma}
p_{\psi\alpha}\epsilon_{\beta\nu\lambda\sigma}p_\psi^\nu \tilde
t^{(1)\lambda}_{(12)} \tilde t^{(1)\sigma}_{(34)} \tilde
T^{(1)}_{((12)(34))\gamma} [f_{(12)}^{K^*}f_{(34)}^{K^{*'}} -
f_{(34)}^{K^*}f_{(12)}^{K^{*'}}],\\
<K^*K^{*'}|12> \!&=&\! P^{(2)\mu\nu\alpha\beta} (p_\psi)\tilde
t^{(1)}_{(12)\alpha} \tilde t^{(1)}_{(34)\beta} \tilde
T^{(1)}_{((12)(34))\nu} [f_{(12)}^{K^*}f_{(34)}^{K^{*'}} +
f_{(34)}^{K^*}f_{(12)}^{K^{*'}}],\\
<K^*K^{*'}|32> \!&=&\! \tilde
T^{(3)\mu\nu\lambda}_{((12)(34))}\tilde t^{(1)}_{(12)\nu} \tilde
t^{(1)}_{(34)\lambda}[f_{(12)}^{K^*}f_{(34)}^{K^{*'}} +
f_{(34)}^{K^*}f_{(12)}^{K^{*'}}].
\end{eqnarray}
Smaller contribution from $KK^*_2$ with $K^*_2\to K^*\pi~or~K\rho$
may needed. The corresponding formulae are
\begin{eqnarray}
<KK^*_2(K^*\pi)|22> \!&=&\! \epsilon^{\mu\nu\lambda\sigma}
p_{\psi\sigma} [\tilde T^{(2)\delta}_{(K^*_23)\nu}
P^{(2)}_{\lambda\delta\alpha\beta}(p_{(124)})
\epsilon^{\alpha\gamma\eta\omega}p_{(124)\gamma} \tilde
t^{(2)\beta}_{(K^*4)\eta} \tilde
t^{(1)}_{(12)\omega}f^{(K^*_2)}_{(124)}f^{(K^*)}_{(12)}
\nonumber\\ & &  - \tilde T^{(2)\delta}_{(\bar K^*_21)\nu}
P^{(2)}_{\lambda\delta\alpha\beta}(p_{(234)})
\epsilon^{\alpha\gamma\eta\omega}p_{(234)\gamma} \tilde
t^{(2)\beta}_{(\bar K^*2)\eta} \tilde
t^{(1)}_{(34)\omega}]f^{(K^*_2)}_{(234)}f^{(K^*)}_{(34)},
\\
<KK^*_{2^+}(K\rho)|22> \!&=&\! \epsilon^{\mu\nu\lambda\sigma}
p_{\psi\sigma} [\tilde T^{(2)\delta}_{(K^*_23)\nu}
P^{(2)}_{\lambda\delta\alpha\beta}(p_{(124)})
\epsilon^{\alpha\gamma\eta\omega}p_{(124)\gamma} \tilde
t^{(2)\beta}_{(\rho 1)\eta} \tilde t^{(1)}_{(24)\omega}
f^{(K^*_2)}_{(124)}f^{(\rho)}_{(24)} \nonumber\\
& &  - \tilde T^{(2)\delta}_{(\bar K^*_21)\nu}
P^{(2)}_{\lambda\delta\alpha\beta}(p_{(234)})
\epsilon^{\alpha\gamma\eta\omega}p_{(234)\gamma} \tilde
t^{(2)\beta}_{(\rho 3)\eta} \tilde t^{(1)}_{(24)\omega}]
f^{(K^*_2)}_{(234)}f^{(\rho)}_{(24)}.
\end{eqnarray}
Some other intermediate states may also need to be considered. The
formulae for $\psi\to KK^*_{2^-}$ with $K^*_{2^-}\to K^*_2
\pi~or~K^*\pi~or~\rho K$ are
\begin{eqnarray}
<KK^*_{2^-}(K^*_2\pi)|12> \!&=&\!
P^{(2)\mu\nu\lambda\sigma}(p_{(124)}) \tilde
T^{(1)}_{(K^*_{2^-}3)\nu}
P^{(2)}_{\lambda\sigma\alpha\beta}(p_{(12)}) \tilde
t^{(2)\alpha\beta}_{(12)}f^{(K^*_{2^-})}_{(124)}f^{(K^*_2)}_{(12)}
\nonumber\\ & &  - P^{(2)\mu\nu\lambda\sigma}(p_{(234)}) \tilde
T^{(1)}_{(\bar K^*_{2^-}1)\nu}
P^{(2)}_{\lambda\sigma\alpha\beta}(p_{(34)}) \tilde
t^{(2)\alpha\beta}_{(34)}f^{(K^*_{2^-})}_{(234)}f^{(K^*_2)}_{(34)}, \\
<KK^*_{2^-}(K^*_2\pi)|32> \!&=&\! \tilde
T^{(3)\mu\nu\gamma}_{(K^*_{2^-}3)}
P^{(2)}_{\nu\gamma\lambda\sigma}(p_{(124)})
P^{(2)\lambda\sigma\alpha\beta}(p_{(12)}) \tilde
t^{(2)}_{(12)\alpha\beta}f^{(K^*_{2^-})}_{(124)}f^{(K^*_2)}_{(12)}
\nonumber\\ & &  - \tilde T^{(3)\mu\nu\gamma}_{(\bar K^*_{2^-}1)}
P^{(2)}_{\nu\gamma\lambda\sigma}(p_{(234)})
P^{(2)\lambda\sigma\alpha\beta}(p_{(34)}) \tilde
t^{(2)}_{(34)\alpha\beta}f^{(K^*_{2^-})}_{(234)}f^{(K^*_2)}_{(34)},\\
<KK^*_{2^-}(K^*\pi)|12> \!&=&\!
P^{(2)\mu\nu\lambda\sigma}(p_{(124)}) \tilde
T^{(1)}_{(K^*_{2^-}3)\nu}
P^{(2)}_{\lambda\sigma\alpha\beta}(p_{(12)}) \tilde
t^{(1)\alpha}_{(K^*4)} \tilde t^{(1)\beta}_{(12)}
f^{(K^*_{2^-})}_{(124)}f^{(K^*)}_{(12)} \nonumber\\ & &  -
P^{(2)\mu\nu\lambda\sigma}(p_{(234)}) \tilde T^{(1)}_{(\bar
K^*_{2^-}1)\nu} P^{(2)}_{\lambda\sigma\alpha\beta}(p_{(34)})
\tilde t^{(1)\alpha}_{(K^*2)} \tilde t^{(1)\beta}_{(34)}
f^{(K^*_{2^-})}_{(234)}f^{(K^*)}_{(34)}, \quad \\
<KK^*_{2^-}(K^*\pi)|32> \!&=&\! \tilde
T^{(3)\mu\nu\gamma}_{(K^*_{2^-}3)}
P^{(2)}_{\nu\gamma\lambda\sigma}(p_{(124)})
P^{(2)\lambda\sigma\alpha\beta}(p_{(12)}) \tilde
t^{(1)}_{(K^*4)\alpha} \tilde t^{(1)}_{(12)\beta}
f^{(K^*_{2^-})}_{(124)}f^{(K^*)}_{(12)} \nonumber\\ & &  - \tilde
T^{(3)\mu\nu\gamma}_{(\bar K^*_{2^-}1)}
P^{(2)}_{\nu\gamma\lambda\sigma}(p_{(234)})
P^{(2)\lambda\sigma\alpha\beta}(p_{(34)}) \tilde
t^{(1)}_{(K^*2)\alpha} \tilde t^{(1)}_{(34)\beta}
f^{(K^*_{2^-})}_{(234)}f^{(K^*)}_{(34)},\\
<KK^*_{2^-}(\rho K)|12> \!&=&\!
P^{(2)\mu\nu\lambda\sigma}(p_{(124)}) \tilde
T^{(1)}_{(K^*_{2^-}3)\nu}
P^{(2)}_{\lambda\sigma\alpha\beta}(p_{(12)}) \tilde
t^{(1)\alpha}_{(\rho 1)} \tilde t^{(1)\beta}_{(24)}
f^{(K^*_{2^-})}_{(124)}f^{(\rho)}_{(24)} \nonumber\\ & &  -
P^{(2)\mu\nu\lambda\sigma}(p_{(234)}) \tilde T^{(1)}_{(\bar
K^*_{2^-}1)\nu} P^{(2)}_{\lambda\sigma\alpha\beta}(p_{(34)})
\tilde t^{(1)\alpha}_{(\rho 3)} \tilde t^{(1)\beta}_{(24)}
f^{(K^*_{2^-})}_{(234)}f^{(\rho)}_{(24)}, \\
<KK^*_{2^-}(\rho K)|32> \!&=&\! \tilde
T^{(3)\mu\nu\gamma}_{(K^*_{2^-}3)}
P^{(2)}_{\nu\gamma\lambda\sigma}(p_{(124)})
P^{(2)\lambda\sigma\alpha\beta}(p_{(12)}) \tilde t^{(1)}_{(\rho
1)\alpha} \tilde t^{(1)}_{(24)\beta}
f^{(K^*_{2^-})}_{(124)}f^{(\rho)}_{(24)} \nonumber\\ & &  - \tilde
T^{(3)\mu\nu\gamma}_{(\bar K^*_{2^-}1)}
P^{(2)}_{\nu\gamma\lambda\sigma}(p_{(234)})
P^{(2)\lambda\sigma\alpha\beta}(p_{(34)}) \tilde t^{(1)}_{(\rho
3)\alpha} \tilde t^{(1)}_{(24)\beta}
f^{(K^*_{2^-})}_{(234)}f^{(\rho)}_{(24)}.
\end{eqnarray}
The formulae for $\psi\to KK'$ with $K'$ the excited $0^{-}$ state
and $K'\to K^*\pi~or~\rho K~or~K^*_0\pi$ are
\begin{eqnarray}
<KK'(K^*\pi)|10> \!&=&\! \tilde T^{(1)\mu}_{(K'3)} \tilde
t^{(1)\nu}_{K^*4}\tilde
t^{(1)}_{(12)\nu}f^{(K')}_{(124)}f^{(K^*)}_{(12)} - \tilde
T^{(1)\mu}_{(\bar K'1)} \tilde t^{(1)\nu}_{\bar K^*2}\tilde
t^{(1)}_{(34)\nu}f^{(K')}_{(234)}f^{(K^*)}_{(34)},\\
<KK'(\rho K)|10> \!&=&\! \tilde T^{(1)\mu}_{(K'3)} \tilde
t^{(1)\nu}_{\rho 1}\tilde
t^{(1)}_{(24)\nu}f^{(K')}_{(124)}f^{(\rho)}_{(24)} - \tilde
T^{(1)\mu}_{(\bar K'1)} \tilde t^{(1)\nu}_{\rho 3}\tilde
t^{(1)}_{(24)\nu}f^{(K')}_{(234)}f^{(\rho)}_{(24)},\\
<KK'(K^*_0\pi)|10> \!&=&\! \tilde T^{(1)\mu}_{(K'3)}
f^{(K')}_{(124)}f^{(K^*_0)}_{(12)} - \tilde T^{(1)\mu}_{(\bar
K'1)}f^{(K')}_{(234)}f^{(K^*_0)}_{(34)}.
\end{eqnarray}

\subsection{$\psi\to\phi\pi^+\pi^-\pi^+\pi^-\to K^+K^-\pi^+\pi^-\pi^+\pi^-$}

As for the $\psi\to\phi\pi^+\pi^-$ channel, the dominant
intermediate states are also $\phi f_0$ and $\phi f_2$. The $f_0$
resonances decay to $\pi^+\pi^-\pi^+\pi^-$ usually through
$\sigma\sigma$ and $\rho\rho$; and $f_2$ resonances decay to
$\pi^+\pi^-\pi^+\pi^-$ usually through $\sigma\sigma$, $\rho\rho$
and $f_2(1270)\sigma$. We assume a similar notation to the
$\psi\to\phi\pi^+\pi^-$ case and number the additional
$\pi^+\pi^-$ as particle 5, 6. Then the corresponding partial wave
amplitudes involving $f_J\to \sigma\sigma$ are:
\begin{eqnarray}
<\phi f_0|01>_{(\sigma\sigma)} \!&=&\! \tilde t^{(1)\mu}_{(12)}
f_{(12)}^{(\phi)}f_{(\sigma\sigma)}^{(f_0)}
\Bigl[f^{(\sigma)}_{(34)}f^{(\sigma)}_{(56)}
+f^{(\sigma)}_{(36)}f^{(\sigma)}_{(45)}\Bigr],\\
<\phi f_0|21>_{(\sigma\sigma)} \!&=&\! \tilde T^{(2)\mu\nu}_{(\phi
f_0)} \tilde t^{(1)}_{(12)\nu} f_{(12)}^{(\phi)}
f_{(\sigma\sigma)}^{(f_0)}
\Bigl[f^{(\sigma)}_{(34)}f^{(\sigma)}_{(56)}
+f^{(\sigma)}_{(36)}f^{(\sigma)}_{(45)}\Bigr], \\
<\phi f_2|01>_{(\sigma\sigma)} \!&=&\!
T^{(f_2)\mu\nu}_{(\sigma\sigma)}
\tilde t^{(1)}_{(12)\nu} f_{(12)}^{(\phi)}f_{(\sigma\sigma)}^{(f_2)} , \\
<\phi f_2|21>_{(\sigma\sigma)} \!&=&\! \tilde
T^{(2)\mu\alpha}_{(\phi f_2)} \tilde
T^{(f_2)}_{(\sigma\sigma)\alpha\nu}\tilde t^{(1)\nu}_{(12)}
f_{(12)}^{(\phi)} f_{(\sigma\sigma)}^{(f_2)}, \\
<\phi f_2|22>_{(\sigma\sigma)} \!&=&\!
\epsilon^{\mu\alpha\beta\gamma} p_{\psi\alpha} \tilde
T^{(2)\delta}_{(\phi f_2)\beta} [\epsilon_{\gamma\lambda\sigma\nu}
\tilde T^{(f_2)\lambda}_{(\sigma\sigma)\delta}
+\epsilon_{\delta\lambda\sigma\nu} \tilde
T^{(f_2)\lambda}_{(\sigma\sigma)\gamma}]p_\psi^\sigma \tilde
t^{(1)\nu}_{(12)}
f_{(12)}^{(\phi)}f_{(\sigma\sigma)}^{(f_2)}, \\
<\phi f_2|23>_{(\sigma\sigma)} \!&=&\!
P^{(3)\mu\alpha\beta\gamma\delta\nu}(p_\psi) \tilde T^{(2)}_{(\phi
f_2)\alpha\beta} \tilde T^{(f_2)}_{(\sigma\sigma)\gamma\delta}
\tilde t^{(1)}_{(12)\nu}
f_{(12)}^{(\phi)}f_{(\sigma\sigma)}^{(f_2)}, \\
<\phi f_2|43>_{(\sigma\sigma)} \!&=&\! \tilde
T^{(4)\mu\nu\lambda\sigma}_{(\phi f_2)}
        \tilde t^{(1)}_{(12)\nu} T^{(f_2)}_{(\sigma\sigma)\lambda\sigma}
        f_{(12)}^{(\phi)}f_{(\sigma\sigma)}^{(f_2)}
\end{eqnarray}
with
\begin{equation}
T^{(f_2)\mu\nu}_{(\sigma\sigma)}=\tilde
t^{(2)\mu\nu}_{(\sigma_{34}\sigma_{56})}
f^{(\sigma)}_{(34)}f^{(\sigma)}_{(56)} + \tilde
t^{(2)\mu\nu}_{(\sigma_{36}\sigma_{45})}
f^{(\sigma)}_{(36)}f^{(\sigma)}_{(45)} .
\end{equation}
For $f_J\to\rho\rho$, if we limit $\rho\rho$ to a relative $l=0$
state, then the corresponding partial wave amplitudes are:
\begin{eqnarray}
<\phi f_0|01>_{(\rho\rho)} \!&=&\! \tilde t^{(1)\mu}_{(12)}
f_{(12)}^{(\phi)}f_{(\rho\rho)}^{(f_0)}
\Bigl[f^{(\rho)}_{(34)}f^{(\rho)}_{(56)}\tilde
t^{(1)\alpha\beta}_{(34)} \tilde t^{(1)}_{(56)\alpha\beta}
+f^{(\rho)}_{(36)}f^{(\rho)}_{(45)}\tilde
t^{(1)\alpha\beta}_{(36)}
\tilde t^{(1)}_{(45)\alpha\beta}\Bigr],\\
<\phi f_0|21>_{(\rho\rho)} \!&=&\! \tilde T^{(2)\mu\nu}_{(\phi
f_0)} \tilde t^{(1)}_{(12)\nu} f_{(12)}^{(\phi)}
f_{(\rho\rho)}^{(f_0)}
\Bigl[f^{(\rho)}_{(34)}f^{(\rho)}_{(56)}\tilde
t^{(1)\alpha\beta}_{(34)} \tilde t^{(1)}_{(56)\alpha\beta}
+f^{(\rho)}_{(36)}f^{(\rho)}_{(45)}\tilde
t^{(1)\alpha\beta}_{(36)}
\tilde t^{(1)}_{(45)\alpha\beta}\Bigr] ,  \\
<\phi f_2|01>_{(\rho\rho)} \!&=&\! T^{(f_2)\mu\nu}_{(\rho\rho)}
\tilde t^{(1)}_{(12)\nu} f_{(12)}^{(\phi)}f_{(\rho\rho)}^{(f_2)} , \\
<\phi f_2|21>_{(\rho\rho)} \!&=&\! \tilde T^{(2)\mu\alpha}_{(\phi
f_2)} \tilde T^{(f_2)}_{(\rho\rho)\alpha\nu}\tilde
t^{(1)\nu}_{(12)}
f_{(12)}^{(\phi)} f_{(\rho\rho)}^{(f_2)}, \\
<\phi f_2|22>_{(\rho\rho)} \!&=&\! \epsilon^{\mu\alpha\beta\gamma}
p_{\psi\alpha} \tilde T^{(2)\delta}_{(\phi f_2)\beta}
[\epsilon_{\gamma\lambda\sigma\nu} \tilde
T^{(f_2)\lambda}_{(\rho\rho)\delta}
+\epsilon_{\delta\lambda\sigma\nu} \tilde
T^{(f_2)\lambda}_{(\rho\rho)\gamma}]p_\psi^\sigma \tilde
t^{(1)\nu}_{(12)}
f_{(12)}^{(\phi)}f_{(\rho\rho)}^{(f_2)}, \\
<\phi f_2|23>_{(\rho\rho)} \!&=&\!
P^{(3)\mu\alpha\beta\gamma\delta\nu}(p_\psi) \tilde T^{(2)}_{(\phi
f_2)\alpha\beta} \tilde T^{(f_2)}_{(\rho\rho)\gamma\delta} \tilde
t^{(1)}_{(12)\nu}f_{(12)}^{(\phi)}f_{(\rho\rho)}^{(f_2)}, \\
<\phi f_2|43>_{(\rho\rho)} \!&=&\! \tilde
T^{(4)\mu\nu\lambda\sigma}_{(\phi f_2)}
        \tilde t^{(1)}_{(12)\nu} T^{(f_2)}_{(\rho\rho)\lambda\sigma}
f_{(12)}^{(\phi)}f_{(\rho\rho)}^{(f_2)}
\end{eqnarray}
where
\begin{eqnarray}
T^{(f_2)\mu\nu}_{(\rho\rho)} &=& P^{(2)\mu\nu\alpha\beta}(p_{f_2})
\Bigl[\tilde t^{(1)}_{(34)\alpha}\tilde t^{(1)}_{(56)\beta}
f^{(\rho)}_{(34)}f^{(\rho)}_{(56)} + \tilde
t^{(1)}_{(36)\alpha}\tilde t^{(1)}_{(45)\beta}
f^{(\rho)}_{(36)}f^{(\rho)}_{(45)}\Bigr].
\end{eqnarray}

For $f_2\to f_2(1270)\sigma$, if we also limit $f_2(1270)\sigma$
to the $l=0$ state, then we have their corresponding partial wave
amplitudes:
\begin{eqnarray}
<\phi f_2|01>_{(f_2\sigma)} \!&=&\! T^{(f_2)\mu\nu}_{(f_2\sigma)}
\tilde t^{(1)}_{(12)\nu} f_{(12)}^{(\phi)}f_{(f_2\sigma)}^{(f_2)} , \\
<\phi f_2|21>_{(f_2\sigma)} \!&=&\! \tilde T^{(2)\mu\alpha}_{(\phi
f_2)} \tilde T^{(f_2)}_{(f_2\sigma)\alpha\nu}\tilde
t^{(1)\nu}_{(12)}f_{(12)}^{(\phi)} f_{(f_2\sigma)}^{(f_2)}, \\
<\phi f_2|22>_{(f_2\sigma)} \!&=&\!
\epsilon^{\mu\alpha\beta\gamma} p_{\psi\alpha} \tilde
T^{(2)\delta}_{(\phi f_2)\beta} [\epsilon_{\gamma\lambda\sigma\nu}
\tilde T^{(f_2)\lambda}_{(f_2\sigma)\delta}
+\epsilon_{\delta\lambda\sigma\nu} \tilde
T^{(f_2)\lambda}_{(f_2\sigma)\gamma}]p_\psi^\sigma \tilde
t^{(1)\nu}_{(12)}
f_{(12)}^{(\phi)}f_{(f_2\sigma)}^{(f_2)}, \\
<\phi f_2|23>_{(f_2\sigma)} \!&=&\!
P^{(3)\mu\alpha\beta\gamma\delta\nu}(p_\psi) \tilde T^{(2)}_{(\phi
f_2)\alpha\beta} \tilde T^{(f_2)}_{(f_2\sigma)\gamma\delta} \tilde
t^{(1)}_{(12)\nu}f_{(12)}^{(\phi)}f_{(f_2\sigma)}^{(f_2)}, \\
<\phi f_2|43>_{(f_2\sigma)} \!&=&\! \tilde
T^{(4)\mu\nu\lambda\sigma}_{(\phi f_2)}
        \tilde t^{(1)}_{(12)\nu} T^{(f_2)}_{(f_2\sigma)\lambda\sigma}
f_{(12)}^{(\phi)}f_{(f_2\sigma)}^{(f_2)}
\end{eqnarray}
with
\begin{eqnarray}
T^{(f_2)\mu\nu}_{(f_2\sigma)} &=&
P^{(2)\mu\nu\alpha\beta}(p_{f_2}) \Bigl[\tilde
t^{(2)}_{(34)\alpha\beta}
f^{(f_2)}_{(34)}f^{(\sigma)}_{(56)}+\tilde
t^{(2)}_{(56)\alpha\beta}
f^{(f_2)}_{(56)}f^{(\sigma)}_{(34)} \nonumber\\
& & \quad\quad\quad\quad+ \tilde t^{(2)}_{(36)\alpha\beta}
f^{(f_2)}_{(36)}f^{(\sigma)}_{(45)}+ \tilde
t^{(2)}_{(45)\alpha\beta}
f^{(f_2)}_{(45)}f^{(\sigma)}_{(36)}\Bigr].
\end{eqnarray}
Unlike the $\psi\to\phi\pi^+\pi^-$ channel, for
$\psi\to\phi\pi^+\pi^-\pi^+\pi^-$ it is possible to go through
$0^{-+}$ resonances ($\eta^*$) decaying to $\pi^+\pi^-\pi^+\pi^-$
through $\rho\rho$. The corresponding partial wave is
\begin{eqnarray}
U^\mu_{\eta^*} \!&=&\! \epsilon^{\mu\nu\alpha\beta}\tilde
t^{(1)}_{(12)\nu} \tilde T^{(1)}_{(\phi\eta^*)\alpha}p_{\psi\beta}
                         \epsilon_{\tau\sigma\gamma\eta}p_3^\tau
                         p_4^\sigma p_5^\gamma p_6^\eta
                         [f_{(34)}^{(\rho)}f_{(56)}^{(\rho)}
                         B_1(Q_{\rho 34})B_1(Q_{\rho 56})
                         B_1(Q_{\eta^*\rho_{34}\rho_{56}})
                         \nonumber \\
                   &   & -f_{(36)}^{(\rho)}f_{(45)}^{(\rho)}
                         B_1(Q_{\rho 36})B_1(Q_{\rho 45})
                         B_1(Q_{\eta^*\rho_{36}\rho_{45}})] .
\end{eqnarray}

Besides partial wave amplitudes given above, for
$\pi^+\pi^-\pi^+\pi^-$ final states, there are many other possible
intermediate states, such as $a_2\pi$, $a_1\pi$, $\pi(1300)\pi$
etc. Before performing partial wave analysis, one should check
various invariant mass spectrum to see what resonances are present
in the data and add the corresponding partial wave amplitudes.

\section{Formalism for $\psi$ radiative decay to mesons}

We denote the $\psi$ polarization four-vector by $\psi_\mu(m_1)$
and the polarization vector of the photon by $e_\nu(m_2)$. Then
the general form for the decay amplitude is
\begin{equation}
A=\psi_\mu(m_1) e^*_\nu(m_2) A^{\mu\nu} =\psi_\mu(m_1)
e^*_\nu(m_2)\sum_i\Lambda_i U_i^{\mu\nu}.
\end{equation}
For the  photon polarization four vector $e_\nu$ with photon
momentum $q$, there is the usual Lorentz orthogonality condition
$e_\nu q^\nu=0$. This is the same as for a massive vector meson.
However, for the photon, there is an additional gauge invariance
condition. Here we assume the Coulomb gauge in the $\psi$ rest
system, {\sl i.e.}, $e_\nu p^\nu_\psi=0$. Then we have
\cite{Greiner}
\begin{equation}
\sum_m e^*_\mu(m) e_\nu(m) = -g_{\mu\nu}+\frac{q_\mu K_\nu+
K_\mu q_\nu}{q\!\cdot\! K}-\frac{K\!\cdot\! K}{(q\!\cdot\! K)^2}q_\mu
q_\nu
\equiv -g^{(\perp\perp)}_{\mu\nu}
\end{equation}
with $K=p_\psi-q$ and $e_\nu K^\nu=0$. The radiative decay cross
section is:
\begin{eqnarray}
\frac{d\sigma}{d\Phi_n}\!&=&\!\frac{1}{2}\sum^2_{m_1=1}\sum^2_{m_2=1}
\psi_{\mu}(m_1) e^*_{\nu}(m_2)A^{\mu\nu}
\psi^*_{\mu'}(m_1) e_{\nu'}(m_2)A^{*\mu'\nu'} \nonumber\\
\!&=&\!-\frac{1}{2}\sum^2_{m_1=1}
\psi_{\mu}(m_1)\psi^*_{\mu'}(m_1) g^{(\perp\perp)}_{\nu\nu'}
A^{\mu\nu}A^{*\mu'\nu'}\nonumber\\
&=& -\frac{1}{2}\sum^2_{\mu=1}A_{\mu\nu}
g^{(\perp\perp)}_{\nu\nu'}A^{*\mu\nu'}\nonumber\\
&=& -\frac{1}{2}\sum_{i,j}\Lambda_i\Lambda_j^*\sum^2_{\mu=1}
U_i^{\mu\nu}g^{(\perp\perp)}_{\nu\nu'}U_j^{*\mu\nu'}
 \equiv \sum_{i,j}P_{ij}\cdot F_{ij}
\end{eqnarray}
where
\begin{eqnarray}
P_{ij} &= P^*_{ji} &= \Lambda_i\Lambda^*_j, \\
F_{ij} &= F^*_{ji} &= -\frac{1}{2}\sum^2_{\mu=1}
U_i^{\mu\nu}g^{(\perp\perp)}_{\nu\nu'}U_j^{*\mu\nu'}.
\end{eqnarray}

Due to the special properties (massless and gauge invariance) of
the photon, the number of independent partial wave amplitudes for
a $\psi$ radiative decay is smaller than for the corresponding
decay to a massive vector meson. For example, for $\psi\to\phi
f_0$, there are two independent partial wave amplitudes with
orbital angular momentum $L=0$ and 2, respectively, which give
different angular distributions; but for $\psi\to\gamma f_0$, with
the gauge invariance condition, the two amplitudes will give the
same angular distribution. So for the $\psi$ radiative decay, the
L-S scheme is not useful any more for choosing independent
amplitudes. One may simply use momenta of the particles to
construct covariant tensor amplitudes;  it is sufficient to check
the helicity amplitudes to make sure there is the right number of
independent amplitudes. From the helicity formalism, it is easy to
show that there is one independent amplitude for $\psi$ radiative
decay to a spin-0 meson, two independent amplitudes for $\psi$
radiative decay to a spin-1 meson, and three independent
amplitudes for  $\psi$ radiative decay to a meson with spin larger
than 1.

\subsection{$\psi$ radiative decay to two pseudoscalar mesons}

We denote the two pseudoscalar mesons as $\pi^+$ and $\pi^-$. For
the decay vertex $\psi\to\gamma f_J$, there are two independent
momenta which we choose to be $p_\psi$ and the momentum of the
photon $q$. We use these two momenta and spin wave functions of
the three particles to construct the covariant tensor amplitudes.

For $\psi\to\gamma f_0$, the $e_\mu$ can only contract with
$\psi^\mu$ since $e_\mu p_\psi^\mu=e_\mu q^\mu=0$; hence there is
only one independent amplitude:
\begin{equation}
U^{\mu\nu}_{\gamma f_0} = g^{\mu\nu} f^{(f_0)}.
\end{equation}

For $\psi\to\gamma f_2$ or $\psi\to\gamma f_4$, the $e_\mu$ may
contract with $\psi^\mu$ or with the spin wave function of $f_J$.
Then $\psi^\mu$ may contract with $e_\mu$, or $q_\mu$, or the spin
wave function of $f_J$; this gives three independent covariant
tensor amplitudes for each $f_J$:
\begin{eqnarray}
U^{\mu\nu}_{(\gamma f_2)1}
\!&=&\! \tilde t^{(f_2)\mu\nu}f^{(f_2)}, \\
U^{\mu\nu}_{(\gamma f_2)2} \!&=&\! g^{\mu\nu}p_\psi^\alpha
p_\psi^\beta \tilde t^{(f_2)}_{\alpha\beta}
B_2(Q_{\Psi\gamma f_2})f^{(f_2)}, \\
U^{\mu\nu}_{(\gamma f_2)3}
 \!&=&\! q^\mu \tilde t^{(f_2)\nu}_{\alpha}p_\psi^\alpha
                B_2(Q_{\psi\gamma f_2})f^{(f_2)}, \\
U^{\mu\nu}_{(\gamma f_4)1}
 \!&=&\! \tilde t^{(f_4)\mu\nu}_{\alpha\beta}p_\psi^\alpha p_\psi^\beta
B_2(Q_{\Psi\gamma f_4})f^{(f_4)} ,\\
U^{\mu\nu}_{(\gamma f_4)2} \!&=&\! g^{\mu\nu} \tilde
t^{(f_4)}_{\alpha\beta\gamma\delta} p_\psi^\alpha p_\psi^\beta
p_\psi^\gamma p_\psi^\delta B_4(Q_{\psi\gamma f_4})f^{(f_4)} ,\\
U^{\mu\nu}_{(\gamma f_4)3} \!&=&\! q^\mu \tilde
t^{(f_4)\nu}_{\alpha\beta\gamma} p_\psi^\alpha p_\psi^\beta
p_\psi^\gamma B_4(Q_{\Psi\gamma f_4})f^{(f_4)}
\end{eqnarray}
where
\begin{eqnarray}
\tilde t^{(f_J)}_{\mu_1\cdots\mu_J}
&=&\sum_m\phi_{\mu_1\cdots\mu_J}(p_{f_J},m)
\phi_{\mu'_1\cdots\mu'_J}(p_{f_J},m) r^{\mu'_1}_\pi\cdots
r^{\mu'_J}_\pi B_J(Q_{f_J\pi\pi}) \nonumber \\ &=&
P_{\mu_1\cdots\mu_J\mu'_1\cdots\mu'_J}(p_{f_J})
r^{\mu'_1}_\pi\cdots r^{\mu'_J}_\pi B_J(Q_{f_J\pi\pi})
\end{eqnarray}
with $J=0$,2; here $r_\pi$ represents the relative momentum
between two pseudoscalar mesons.

We use $p_\psi$ instead of $q$ to contract with $\tilde t^{(f_J)}$
because $q\tilde t^{(f_J)}=p_\psi \tilde t^{(f_J)}$ and $p_\psi$
has only a time component in the $\psi$ rest system. This makes
the calculation simpler.

\subsection{$\psi\to\gamma\eta\pi^+\pi^-$}

This is a three step process: $\psi\to\gamma X$ with $X\to yz$ and
$y\to\pi\pi$ or $y\to\eta\pi$. The amplitudes $U^i_{\mu\nu}$ are
listed using the notation:
\begin{equation}
<\gamma J^{PC}|(yz)i> \nonumber
\end{equation}
where J,P,C are the intrinsic spin, parity and C-parity of the X
particle, respectively. We denote $\pi^+$, $\pi^-$, $\eta$ as
1,2,3, respectively. The possible $J^{PC}$ for $X$ are $0^{-+}$,
$1^{++}$, $1^{-+}$, $2^{++}$, $2^{-+}$, $3^{++}$, $3^{-+}$, etc.
For invariant mass below 2 GeV, we consider $J$ up to 2. For
$\psi\to\gamma X$, we choose two independent momenta $p_\psi$ for
$\psi$ and $q$ for the photon to be contracted with spin wave
functions.

For the $\psi\to\gamma 0^{-+}$ vertex, there is only one
independent coupling, $\epsilon_{\mu\nu\lambda\sigma}\psi^\mu
e^\nu q^\lambda p^\sigma_\psi$. With various possible $yz$ states,
we have $U^i_{\mu\nu}$ for $\psi\to\gamma 0^-\to\eta\pi^+\pi^-$ as
follows:
\begin{eqnarray}
<\gamma 0^{-+}|(f_0\eta)1> \!&=&\! S_{\mu\nu}
                          B_1(Q_{\psi\gamma X})f_{(12)}^{(f_0)},\\
<\gamma 0^{-+}|(a_0\pi)1> \!&=&\! S_{\mu\nu}
                          B_1(Q_{\psi\gamma X})
                         (f_{(13)}^{(a_0)}+f_{(23)}^{(a_0)}),\\
<\gamma 0^{-+}|(f_2\eta)1> \!&=&\! S_{\mu\nu} B_1(Q_{\psi\gamma
X})f_{(12)}^{(f_2)}
\tilde t^{(2)}_{(f_2\eta)\gamma\delta}\tilde t_{(12)}^{(2)\gamma\delta}, \\
<\gamma 0^{-+}|(a_2\pi)1> \!&=&\! S_{\mu\nu} B_1(Q_{\psi\gamma X})
\Bigl\{f_{(13)}^{(a_2)} \tilde t^{(2)}_{(a_22)\gamma\delta}\tilde
t_{(13)}^{(2)\gamma\delta} +f_{(23)}^{(a_2)} \tilde
t^{(2)}_{(a_21)\gamma\delta}\tilde
                t_{(23)}^{(2)\gamma\delta}\Bigr\}
\end{eqnarray}
with $S_{\mu\nu}$ defined as
\begin{equation}
\label{smunu}
S_{\mu\nu}=\epsilon_{\mu\nu\alpha\beta}p_\psi^{\alpha}q^\beta .
\end{equation}

For the $\psi\to\gamma 1^{++}$ vertex, there are two independent
couplings for each $yz$.
\begin{eqnarray}
<\gamma 1^{++}|(f_0\eta)1> \!&=&\!
\epsilon_{\mu\nu\alpha\beta}p_\psi^{\alpha} \tilde
t^{(1)\beta}_{(\eta f_0)}f_{(12)}^{(f_0)}, \\
<\gamma 1^{++}|(a_0\pi)1> \!&=&\!
\epsilon_{\mu\nu\alpha\beta}p_\psi^{\alpha} (\tilde
t^{(1)\beta}_{(a_01)} f_{(23)}^{(a_0)} + \tilde
t^{(1)\beta}_{(a_02)} f_{(13)}^{(a_0)}), \\
<\gamma 1^{++}|(f_0\eta)2> \!&=&\! q_\mu S_{\nu\beta}
\tilde t^{(1)\beta}_{(\eta f_0)} B_2(Q_{\psi\gamma X})f_{(12)}^{(f_0)}, \\
<\gamma 1^{++}|(a_0\pi)2> \!&=&\!  q_\mu S_{\nu\beta}
B_2(Q_{\psi\gamma X}) [t^{(1)\beta}_{(a_01)}f_{(23)}^{(a_0)}
+t^{(1)\beta}_{(a_02)}f_{(13)}^{(a_0)}],  \\
<\gamma 1^{++}|(f_2\eta)1> \!&=&\!
\epsilon_{\mu\nu\alpha\beta}p_\psi^{\alpha}
\tilde{g}_X^{\beta\gamma}\tilde t^{(2)}_{(12)\gamma\delta}
\tilde t^{(1)\delta}_{(\eta f_2)} f_{(12)}^{(f_2)}, \\
<\gamma 1^{++}|(a_2\pi)1> \!&=&\!
\epsilon_{\mu\nu\alpha\beta}p_\psi^{\alpha}\tilde{g}_X^{\beta\gamma}
[\tilde t^{(2)}_{(13)\gamma\delta} \tilde t^{(1)\delta}_{(a_22)}
f_{(13)}^{(a_2)} +\tilde t^{(2)}_{(23)\gamma\delta}\tilde
t^{(1)\delta}_{(a_21)}
f_{(23)}^{(a_2)}] \\
<\gamma 1^{++}|(f_2\eta)2> \!&=&\! q_\mu S_{\nu\beta}
\tilde{g}_X^{\beta\beta'}\tilde t_{(12)\beta'\alpha'} \tilde
t^{(1)\alpha'}_{(\eta f_2)}B_2(Q_{\psi\gamma X})
f_{(12)}^{(f_2)},\\
<\gamma 1^{++}|(a_2\pi)2> \!&=&\! q_\mu S_{\nu\beta}
\tilde{g}_X^{\beta\beta'}B_2(Q_{\psi\gamma X}) [\tilde
t^{(2)}_{(13)\beta'\alpha'}\tilde t^{(1)\alpha'}_{(a_22)}
f_{(13)}^{(a_2)}+\tilde t^{(2)}_{(23)\beta'\alpha'}\tilde
t^{(1)\alpha'}_{(a_21)} f_{(23)}^{(a_2)}] \nonumber\\
\end{eqnarray}
where $\tilde g_X^{\alpha\beta}=g^{\alpha\beta}-\frac{p_X^\alpha
p_X^\beta}{p_X^2}$. For $1^{++}$ decaying to $f_2\eta$ and
$a_2\pi$, the orbital angular momentum $l$ could be 1 and 3; but
we ignore the $l=3$ contribution because of the strong centrifugal
barrier.

For $\psi\to\gamma 1^{-+}$, the exotic $1^{-+}$ meson cannot decay
into $f_0\eta$ and $a_0\pi$. We have four $U^i_{\mu\nu}$
amplitudes here:
\begin{eqnarray}
<\gamma 1^{-+}|(f_2\eta)1> \!&=&\! g_{\mu\nu} S_{\gamma\delta}
\tilde t^{(2)\gamma\sigma}_{(\eta f_2)} \tilde
t^{(2)\delta}_{(12)\sigma}
f_{(12)}^{(f_2)}B_1(Q_{\psi\gamma X}), \\
<\gamma 1^{-+}|(a_2\pi)1> \!&=&\! g_{\mu\nu} S_{\gamma\delta}
[\tilde t^{(2)\gamma\sigma}_{(a_22)}\tilde
t_{(13)\sigma}^{(2)\delta} f_{(13)}^{(a_2)} + \tilde
t^{(2)\gamma\sigma}_{(a_21)}\tilde
t_{(23)\sigma}^{(2)\delta}f_{(23)}^{(a_2)}]B_1(Q_{\psi\gamma X}), \\
<\gamma 1^{-+}|(f_2\eta)2> \!&=&\! q_\mu
\epsilon_{\nu\beta\gamma\delta}K^\beta \tilde
t^{(2)\gamma\sigma}_{(\eta f_2)} \tilde t^{(2)\delta}_{(12)\sigma}
f_{(12)}^{(f_2)}B_1(Q_{\psi\gamma X}), \\
<\gamma 1^{-+}|(a_2\pi)2> \!&=&\! q_\mu
\epsilon_{\nu\beta\gamma\delta}K^\beta [\tilde
t^{(2)\gamma\sigma}_{(a_22)}\tilde t_{(13)\sigma}^{(2)\delta}
f_{(13)}^{(a_2)} + \tilde t^{(2)\gamma\sigma}_{(a_21)}\tilde
t_{(23)\sigma}^{(2)\delta}f_{(23)}^{(a_2)}]B_1(Q_{\psi\gamma X}).
\end{eqnarray}

 For $\psi\to\gamma 2^{++}$, there are three independent
couplings and two possible $yz$ states, $f_2\eta$ and $a_2\pi$.
\begin{eqnarray}
<\gamma 2^{++}|(f_2\eta)1> \!&=&\!
P^{(2)}_{\mu\nu\alpha\lambda}(K)
\epsilon^{\alpha\beta\gamma\delta}K_{\beta} \tilde t^{(1)}_{(\eta
f_2)\gamma}\tilde t_{(12)\delta}^{(2)\lambda}f_{(12)}^{(f_2)}\\
<\gamma 2^{++}|(a_2\pi)1> \!&=&\! P^{(2)}_{\mu\nu\alpha\lambda}(K)
\epsilon^{\alpha\beta\gamma\delta}K_\beta [\tilde
t^{(1)}_{(a_21)\gamma}\tilde t_{(23)\delta}^{(2)\lambda}
f_{(23)}^{(a_2)} + \tilde t^{(1)}_{(a_22)\gamma}\tilde
t_{(13)\delta}^{(2)\lambda} f_{(13)}^{(a_2)}],\\
<\gamma 2^{++}|(f_2\eta)2> \!&=&\! g_{\mu\nu}p_\psi^\lambda
p_\psi^\sigma P^{(2)}_{\lambda\sigma\alpha\beta}(K)
B_2(Q_{\psi\gamma X}) \epsilon^\alpha_{\gamma\delta\alpha'}
K^\gamma \tilde t^{(1)\delta}_{(\eta f_2)}\tilde
t_{(12)}^{(2)\alpha'\beta}f_{(12)}^{(f_2)},   \\
<\gamma 2^{++}|(a_2\pi)2> \!&=&\! g_{\mu\nu}p_\psi^\lambda
p_\psi^\sigma P^{(2)}_{\lambda\sigma\alpha\beta}(K)
\epsilon^\alpha_{\gamma\delta\alpha'} K^\gamma [\tilde
t^{(1)}_{(a_21)\delta}\tilde
t_{(23)}^{(2)\alpha'\lambda} f_{(23)}^{(a_2)} \nonumber\\
&&\quad\quad +\tilde t^{(1)}_{(a_22)\delta}\tilde
t_{(13)}^{(2)\alpha'\lambda}f_{(13)}^{(a_2)}]B_2(Q_{\psi\gamma X}),\\
<\gamma 2^{++}|(f_2\eta)3> \!&=&\! q_\mu
p_\psi^{\lambda}P^{(2)}_{\nu\lambda\alpha\beta}(K)
B_2(Q_{\Psi\gamma X})
\epsilon^\alpha_{\gamma\delta\alpha'}K^\gamma \tilde
t^{(1)\delta}_{(\eta f_2)}\tilde
t_{(12)}^{(2)\alpha'\beta}f_{(12)}^{(f_2)},\\
<\gamma 2^{++}|(a_2\pi)3> \!&=&\!   q_\mu
p_\psi^{\lambda}P^{(2)}_{\nu\lambda\alpha\beta}(K)
\epsilon^\alpha_{\gamma\delta\alpha'}K^\gamma [\tilde
t^{(1)}_{(a_21)\delta}\tilde t_{(23)}^{(2)\alpha'\beta}
f_{(23)}^{(a_2)} 
+\tilde t^{(1)}_{(a_22)\delta}\tilde
t_{(13)}^{(2)\alpha'\beta}f_{(13)}^{(a_2)}]B_2(Q_{\psi\gamma X}).
\nonumber\\
\end{eqnarray}

For $\psi\to\gamma 2^{-+}$, we have
\begin{eqnarray}
<\gamma 2^{-+}|(f_0\eta)1> \!&=&\!
\epsilon_{\mu\nu\alpha\beta}p_\psi^{\alpha} \tilde
t_{(f_0\eta)}^{(2)\beta\gamma}q_\gamma
f_{(12)}^{(f_0)}B_1(Q_{\psi\gamma X}), \\
<\gamma 2^{-+}|(a_0\pi)1> \!&=&\!
\epsilon_{\mu\nu\alpha\beta}p_\psi^{\alpha} \Bigl\{\tilde
t_{(a_01)}^{(2)\beta\gamma}f_{(23)}^{(a_0)}+\tilde
t_{(a_02)}^{(2)\beta\gamma}f_{(13)}^{(a_0)}\Bigr\}
q_\gamma B_1(Q_{\psi\gamma X}), \\
<\gamma 2^{-+}|(f_0\eta)2> \!&=&\! S_{\mu\nu} p_{\psi\gamma}
p_{\psi\delta} \tilde t_{(f_0\eta)}^{(2)\gamma\delta}
f_{(12)}^{(f_0)}B_3(Q_{\psi\gamma X}), \\
<\gamma 2^{-+}|(a_0\pi)2> \!&=&\!  S_{\mu\nu} p_{\psi\gamma}
p_{\psi\delta} \Bigl\{\tilde
t_{(a_01)}^{(2)\gamma\delta}f_{(23)}^{(a_0)}+\tilde
t_{(a_02)}^{(2)\gamma\delta}f_{(13)}^{(a_0)}\Bigr\}
B_3(Q_{\psi\gamma X}), \\
<\gamma 2^{-+}|(f_0\eta)3> \!&=&\! q_\mu S_{\nu\gamma}
\tilde{t}^{(2)\gamma\delta}_{(f_0\eta)}p_{\psi\delta}
f_{(12)}^{(f_0)}B_3(Q_{\psi\gamma X}), \\
<\gamma 2^{-+}|(a_0\pi)3> \!&=&\! q_\mu S_{\nu\gamma}
p_{\psi\delta} \Bigl\{\tilde
t_{(a_01)}^{(2)\gamma\delta}f_{(23)}^{(a_0)}+\tilde
t_{(a_02)}^{(2)\gamma\delta}f_{(13)}^{(a_0)}\Bigr\}
B_3(Q_{\Psi\gamma X}), \\
<\gamma 2^{-+}|(f_2\eta)1> \!&=&\!
\epsilon_{\mu\nu\alpha\beta}p_\psi^{\alpha}q_\gamma
P^{(2)\beta\gamma\beta'\gamma'}(K)\tilde
t^{(2)}_{(12)\beta'\gamma'}
f_{(12)}^{(f_2)} B_1(Q_{\psi\gamma X})\\
<\gamma 2^{-+}|(a_2\pi)1> \!&=&\!
\epsilon_{\mu\nu\alpha\beta}p_\psi^{\alpha}
P^{(2)\beta\gamma\beta'\gamma'}(K)q_\gamma B_1(Q_{\psi\gamma X})
[\tilde t^{(2)}_{(23)\beta'\gamma'}f_{(23)}^{(a_2)}
+\tilde t^{(2)}_{(13)\beta'\gamma'}f_{(13)}^{(a_2)}],
\nonumber\\ & &\\
<\gamma 2^{-+}|(f_2\eta)2> \!&=&\! S_{\mu\nu} p_\psi^\gamma
p_\psi^\delta P^{(2)}_{\gamma\delta\gamma'\delta'}(K)\tilde
t_{(12)}^{(2)\gamma'\delta'}
f_{(12)}^{(f_2)}B_3(Q_{\psi\gamma X}), \\
<\gamma 2^{-+}|(a_2\pi)2> \!&=&\! S_{\mu\nu} p_\psi^\gamma
p_\psi^\delta P^{(2)}_{\gamma\delta\gamma'\delta'}(K)
B_3(Q_{\psi\gamma X})[\tilde t_{(23)}^{(2)\gamma'\delta'}
f_{(23)}^{(a_2)}
+\tilde t_{(13)}^{(2)\gamma'\delta'}f_{(13)}^{(a_2)}], \\
<\gamma 2^{-+}|(f_2\eta)3> \!&=&\! q_\mu S_{\nu\gamma}
p_{\psi\delta} P^{(2)\gamma\delta\gamma'\delta'}(K)\tilde
t^{(2)}_{(12)\gamma'\delta'}
f_{(12)}^{(f_2)}B_3(Q_{\Psi\gamma X}),\\
<\gamma 2^{-+}|(a_2\pi)3> \!&=&\! q_\mu S_{\nu\gamma}
p_{\psi\delta}P^{(2)\gamma\delta\gamma'\delta'}(K)
B_3(Q_{\Psi\gamma X}) [\tilde
t^{(2)}_{(23)\gamma'\delta'}f_{(23)}^{(a_2)} +\tilde
t^{(2)}_{(13)\gamma'\delta'}f_{(13)}^{(a_2)}]
\end{eqnarray}
with $S_{\mu\nu}$ defined as in Eq.(\ref{smunu}).

\subsection{$\psi\to\gamma K\bar K\pi$}

Possible intermediate channels for this process are
$K^*K$,$K_0^*K$,$K^*_2K$,$a_0\pi$,$a_2\pi$. The formulae for
$K_0^*K$,$K^*_2K$,$a_0\pi$,$a_2\pi$ intermediate states to the
$K\bar K\pi$ final state are the same as for the $a_0\pi$,
$a_2\pi$, $f_0\eta$, $f_2\eta$ intermediate states given in the
previous subsection for the $\pi^+\pi^-\eta$ final state. So here
we only give partial wave amplitudes $U^i_{\mu\nu}$ with $K^*K$
intermediate states. We denote $K$,$\bar K$, $\pi$ as particle
1,2,3.

\begin{eqnarray}
<\gamma 0^{-+}|(K^*K)1> \!&=&\! S_{\mu\nu} B_1(Q_{\psi\gamma X})
[\tilde t^{(1)}_{(K^*\bar K)\lambda} \tilde
t^{(1)\lambda}_{(13)}f_{(13)}^{(K^*)}+\tilde t^{(2)}_{(K^*\bar
K)\lambda} \tilde t^{(1)\lambda}_{(23)}f_{(23)}^{(K^*)}],\\
<\gamma 1^{++}|(K^*K)1> \!&=&\!
\epsilon_{\mu\nu\alpha\beta}p_\psi^{\alpha} [\tilde
t_{(23)}^{(1)\beta}f_{(23)}^{(K^*)}+\tilde
t_{(13)}^{(1)\beta}f_{(13)}^{(K^*)}], \\
<\gamma 1^{++}|(K^*K)2> \!&=&\! q_\mu S_{\nu\beta}
B_2(Q_{\psi\gamma X}) [\tilde t_{(23)}^{(1)\beta}
f_{(23)}^{(K^*)}+\tilde t_{(13)}^{(1)\beta}
f_{(13)}^{(K^*)}], \\
<\gamma 1^{-+}|(K^*K)1> \!&=&\! g_{\mu\nu}S_{\gamma\delta} [\tilde
t^{(1)\gamma}_{(K^*\bar K)}\tilde
t^{(1)\delta}_{(13)}f_{(13)}^{(K^*)}+\tilde t^{(1)\gamma}_{(\bar
K^*K )} \tilde
t^{(1)\delta}_{(23)}f_{(23)}^{(K^*)}]B_1(Q_{\psi\gamma X}),
\\
<\gamma 1^{-+}|(K^*K)2> \!&=&\! q_\mu
\epsilon_{\nu\beta\gamma\delta}K^\beta [\tilde
t^{(1)\gamma}_{(K^*\bar K)}\tilde
t^{(1)\delta}_{(13)}f_{(13)}^{(K^*)}+\tilde t^{(1)\gamma}_{(\bar
K^*K )} \tilde
t^{(1)\delta}_{(23)}f_{(23)}^{(K^*)}]B_1(Q_{\psi\gamma X}).
\end{eqnarray}

\subsection{$\psi\to\gamma \pi^+\pi^-\pi^+\pi^-$}

Listed here are formulae used in Refs.\cite{BES4,Scott}.
\begin{eqnarray}
<\gamma 0^{-+}|\rho\rho> \!&=&\! S_{\mu\nu}
\epsilon_{\gamma\delta\lambda\sigma}p_1^\gamma p_2^\delta
p_3^\lambda p_4^\sigma B_1(Q_{\psi\gamma X})
[f_{(12)}^{(\rho)}f_{(34)}^{(\rho)} B_1(Q_{\rho 12}) B_1(Q_{\rho
34}) \nonumber \\
&   &  B_1(Q_{X(12)(34)}) - f_{(14)}^{(\rho)}f_{(32)}^{(\rho)}
B_1(Q_{\rho 14})
B_1(Q_{\rho 32})B_1(Q_{X(14)(32)})], \\
<\gamma 0^{++}|\sigma\sigma>
                  \!&=&\!g_{\mu\nu}[f_{(12)}^{(\sigma)}
                         f_{(34)}^{(\sigma)} +
                         f_{(14)}^{(\sigma)}f_{(32)}^{(\sigma)}], \\
<\gamma 0^{++}|\rho\rho>  \!&=&\! g_{\mu\nu}[f_{(12)}^{(\rho)}
                         f_{(34)}^{(\rho)}(q_{(12)}\!\cdot\! q_{(34)})
                         B_1(Q_{\rho 12})B_1(Q_{\rho 34}) + \nonumber \\
                   &   & f_{(14)}^{(\rho)}f_{(32)}^{(\rho)}
                              (q_{(14)}\!\cdot\! q_{(32)})
                         B_1(Q_{\rho 14})B_1(Q_{\rho 32})], \\
<\gamma 0^{++}|\pi\pi^{\prime}(\pi\sigma)>
                  \!&=&\! g_{\mu\nu}
                         [f_{(123)}^{(\pi^{\prime})}
                         (f_{(12)}^{(\sigma)} + f_{(32)}^{(\sigma)}) +
                          f_{(234)}^{(\pi^{\prime})}
                         (f_{(23)}^{(\sigma)} + f_{(34)}^{(\sigma)}) +
                         \nonumber \\
                   &   & f_{(143)}^{(\pi^{\prime})}
                         (f_{(14)}^{(\sigma)} + f_{(34)}^{(\sigma)}) +
                          f_{(214)}^{(\pi^{\prime})}
                         (f_{(21)}^{(\sigma)} +
                         f_{(14)}^{(\sigma)})],
                         \\
<\gamma 0^{++}|\pi\pi^{\prime}(\pi\rho)>
                  \!&=&\!g_{\mu\nu}[f_{(123)}^{(\pi')}f_{(12)}^{(\rho)}
                          q_{(12)\alpha}(p_3-p_{(12)})^{\alpha}
                         B_1(Q_{\pi'\rho 3})B_1(Q_{\rho 12}) +
                         \nonumber \\
                   &   &  f_{(234)}^{(\pi')}f_{(23)}^{(\rho)}
                          q_{(23)\gamma}(p_4-p_{(23)})^{\gamma}
                         B_1(Q_{\pi'\rho 4})B_1(Q_{\rho 23}) +
                         \nonumber \\ & &
+\{1\leftrightarrow 3\}+\{2\leftrightarrow 4\}
+\{1\leftrightarrow 3 ~\&~ 2\leftrightarrow 4\}], \\
<\gamma 0^{++}|\pi a_{1}(\pi\rho)> \!&=&\!g_{\mu\nu}
[P^{(1)}_{\alpha\beta}(p_{(123)}) p_4^\alpha q_{(12)}^\beta
f_{(123)}^{(a_1)}f_{(12)}^{(\rho)} B_1(Q_{Xa_14})B_1(Q_{\rho 12})
\nonumber\\
& & +P^{(1)}_{\alpha\beta}(p_{(234)}) p_1^\alpha q_{(23)}^\beta
f_{(234)}^{(a_1)}f_{(23)}^{(\rho)} B_1(Q_{Xa_11})B_1(Q_{\rho 23})
                         \nonumber \\ & &
+\{1\leftrightarrow 3\}+\{2\leftrightarrow 4\}
+\{1\leftrightarrow 3 ~\&~ 2\leftrightarrow 4\}],\\
<\gamma 2^{++}|(yy)1>
                  \!&=&\!X_{\mu\nu}^{(yy)},
                         \\
<\gamma 2^{++}|(yy)2>
                  \!&=&\!g_{\mu\nu}p_\psi^\alpha p_\psi^\beta
                  X_{\alpha\beta}^{(yy)}B_2(Q_{\psi X \gamma}),
                          \\
<\gamma 2^{++}|(yy)3> \!&=&\! q_\mu
X_{\nu\alpha}^{yy}p_\psi^\alpha
B_2(Q_{\psi X \gamma}),\\
<\gamma 2^{++}|(f_2\sigma)1> \!&=&\!
P^{(2)}_{\mu\nu\alpha\beta}(K)\tilde t_{(12)}^{\alpha\beta}
f_{(12)}^{(f_2)}f_{(34)}^{(\sigma)} +\{1\leftrightarrow 3\}
+\{2\leftrightarrow 4\} +\{1\leftrightarrow 3 ~\&~
2\leftrightarrow 4\},
                        \nonumber\\ &&\\
<\gamma 2^{++}|(f_2\sigma)2> \!&=&\! g_{\mu\nu} p_\psi^\alpha
p_\psi^\beta P^{(2)}_{\alpha\beta\gamma\delta}(K)\tilde
t_{(12)}^{(2)\gamma\delta}
f_{(12)}^{(f_2)}f_{(34)}^{(\sigma)}B_2(Q_{\psi X \gamma})
+\{1\leftrightarrow 3\} \nonumber\\
& & \quad\quad +\{2\leftrightarrow 4\}+\{1\leftrightarrow 3 ~\&~
2\leftrightarrow 4\}, \\
<\gamma 2^{++}|(f_2\sigma)3> \!&=&\! q_\mu p_\psi^\beta
P^{(2)}_{\nu\beta\gamma\delta}(K)\tilde t_{(12)}^{(2)\gamma\delta}
f_{(12)}^{(f_2)}f_{(34)}^{(\sigma)}B_2(Q_{\psi X \gamma})
+\{1\leftrightarrow 3\} \nonumber\\
& & \quad\quad +\{2\leftrightarrow 4\}+\{1\leftrightarrow 3 ~\&~
2\leftrightarrow 4\}.
\end{eqnarray}

The amplitudes involving X particles of $J^{P}=2^{++}$ involves a
rank two tensor, $X^{(yy)}$. The definition of this is given
below:
\begin{eqnarray}
X_{\mu\nu}^{(\sigma\sigma)}\!&=&\!f_{(12)}^{(\sigma)}f_{(34)}^{(\sigma)}
B_2(Q_{X(12)(34)}) P^{(2)}_{\mu\nu\alpha\beta}(K)
(p_{(12)}^{\alpha}-p_{(34)}^{\alpha})
(p_{(12)}^{\beta}-p_{(34)}^{\beta})+\{2\leftrightarrow 4\}\\
X_{\mu\nu}^{(\rho\rho)} \!&=&\!f_{(12)}^{(\rho)}f_{(34)}^{(\rho)}
B_1(Q_{\rho 12})B_1(Q_{\rho 34})P^{(2)}_{\mu\nu\alpha\beta}(K)
\tilde t_{(12)}^{(1)\alpha} \tilde t_{(34)}^{(1)\beta}
+\{2\leftrightarrow 4\}
\end{eqnarray}
where $L=2$ decay for $X\to\rho\rho$ is ignored in view of the
centrifugal barrier suppression.

From the flux tube model for hybrids, $1^{-+}$ hybrids with $I=0$
decay dominantly into $4\pi$ through $a_1\pi$. Then the
$\psi\to\gamma\pi^+\pi^-\pi^+\pi^-$ is an ideal place for finding
$1^{-+}$ hybrids. With high statistics data at BES and CLEO-C, one
should look for the iso-scalar $1^{-+}$ hybrid in this channel.
Here we add the formulae for $1^{-+}$ hybrid production.
\begin{eqnarray}
<\gamma 1^{-+}|[\pi a_1(\rho\pi)]1> \!&=&\!
g_{\mu\nu}p_\psi^\alpha P^{(1)}_{\alpha\beta}(K)
[P^{(1)\beta\gamma}(p_{(123)}) \tilde
t^{(1)}_{(12)\gamma}f^{(a_1)}_{(123)}f^{(\rho)}_{(12)} \nonumber\\
& & +P^{(1)\beta\gamma}(p_{(234)}) \tilde
t^{(1)}_{(23)\gamma}f^{(a_1)}_{(234)}f^{(\rho)}_{(23)}]
 +\{1\leftrightarrow 3\}+\{2\leftrightarrow 4\}\nonumber\\& &
+\{1\leftrightarrow 3 ~\&~ 2\leftrightarrow 4\},\\
<\gamma 1^{-+}|[\pi a_1(\rho\pi)]2> \!&=&\! q_\mu
P^{(1)}_{\nu\beta}(K) [P^{(1)\beta\gamma}(p_{(123)}) \tilde
t^{(1)}_{(12)\gamma}f^{(a_1)}_{(123)}f^{(\rho)}_{(12)} \nonumber\\
& & +P^{(1)\beta\gamma}(p_{(234)}) \tilde
t^{(1)}_{(23)\gamma}f^{(a_1)}_{(234)}f^{(\rho)}_{(23)}]
 +\{1\leftrightarrow 3\}+\{2\leftrightarrow 4\}\nonumber\\& &
+\{1\leftrightarrow 3 ~\&~ 2\leftrightarrow 4\}.
\end{eqnarray}

\subsection{$\psi\to\gamma K^+K^-\pi^+\pi^-$ }

We construct the amplitudes $U^i_{\mu\nu}$ with a notation similar
to the previous subsection for the
$\psi\to\gamma\pi^+\pi^-\pi^+\pi^-$ channel. Here we denote
$K^+$,$K^-$,$\pi^+$,$\pi^-$ as 1,2,3,4.
\begin{eqnarray}
<\gamma 0^{-+}|K^*\bar K^*> \!&=&\! S_{\mu\nu}
\epsilon_{\gamma\delta\lambda\sigma}p_1^\gamma p_2^\delta
p_3^\lambda p_4^\sigma f_{(14)}^{(K^*)}f_{(23)}^{(\bar K^*)}
\nonumber \\  &  & B_1(Q_{\psi\gamma X)}
B_1(Q_{K^* 14})B_1(Q_{\bar K^*23})B_1(Q_{X(14)(23)}), \\
<\gamma 0^{++}|\kappa\kappa> \!&=&\!
g_{\mu\nu}f_{(14)}^{(\kappa)}f_{(23)}^{(\kappa)},
                \\
<\gamma 0^{++}|K^*\bar K^* > \!&=&\! g_{\mu\nu}
\tilde{t}^{(1)\alpha}_{(14)} \tilde{t}^{(1)}_{(23)\alpha}
f_{(14)}^{(K^*)}f_{(23)}^{(K^*)},
                \\
<\gamma 0^{++}|K^*\kappa> \!&=&\! g_{\mu\nu} [\tilde
t^{(1)}_{(K^*\bar\kappa)\alpha}\tilde{t}_{(14)}^{(1)\alpha}
f_{(14)}^{(K^*)}f_{(23)}^{(\kappa)} +\tilde t^{(1)}_{(\bar
K^*\kappa)\alpha} \tilde{t}_{(23)}^{(1)\alpha}
f_{(23)}^{(K^*)}f_{(14)}^{(\kappa)}],
                \\
<\gamma 1^{++}|(K^*\bar K^*)1> \!&=&\!
\epsilon_{\mu\nu\lambda\alpha}p_\psi^\lambda
\epsilon^{\alpha\beta\gamma\delta}
K_{\beta}\tilde{t}^{(1)}_{(14)\gamma}\tilde{t}^{(1)}_{(23)\delta}
                f_{(14)}^{(K^*)}f_{(23)}^{(\bar K^*)}, \\
<\gamma 1^{++}|(K^*\bar K^*)2> \!&=&\! q_\mu S_{\nu\alpha}
\epsilon^{\alpha\beta\gamma\delta}
K_{\beta}\tilde{t}^{(1)}_{(14)\gamma}\tilde{t}^{(1)}_{(23)\delta}
f_{(14)}^{(K^*)}f_{(23)}^{(\bar K^*)}B_2(Q_{\psi\gamma X)}, \\
<\gamma 1^{++}|(K^*\kappa)1> \!&=&\!
\epsilon_{\mu\nu\lambda\alpha}p_\psi^\lambda
\epsilon^{\alpha\beta\gamma\delta} p_{X\beta} [\tilde
t^{(1)}_{(K^*\bar\kappa)\gamma}\tilde t^{(1)}_{(14)\delta}
f_{(14)}^{(K^*)}f_{(23)}^{(\kappa)} \nonumber\\
& &   +\tilde t^{(1)}_{(\bar K^*\kappa)\gamma}\tilde
t^{(1)}_{(23)\delta}
f_{(23)}^{(K^*)}f_{(14)}^{(\kappa)}], \\
<\gamma 1^{++}|(K^*\kappa)2> \!&=&\! q_\mu S_{\nu\alpha}
\epsilon^{\alpha\beta\gamma\delta} p_{X\beta}[\tilde
t^{(1)}_{(K^*\bar\kappa)\gamma}\tilde t^{(1)}_{(14)\delta}
f_{(14)}^{(K^*)}f_{(23)}^{(\kappa)} \nonumber\\
& &   +\tilde t^{(1)}_{(\bar K^*\kappa)\gamma}\tilde
t^{(1)}_{(23)\delta}
f_{(23)}^{(K^*)}f_{(14)}^{(\kappa)}]B_2(Q_{\psi\gamma X)}, \\
<\gamma 2^{++}|(yz)1>  \!&=&\! X_{\mu\nu}^{(yz)},
                \\
<\gamma 2^{++}|(yz)2>  \!&=&\!g_{\mu\nu}p_\psi^\alpha p_\psi^\beta
X_{\alpha\beta}^{(yz)}B_2(Q_{\psi\gamma X)},
                \\
<\gamma 2^{++}|(yz)3> \!&=&\! q_\mu p_\psi^\alpha
X^{(yz)}_{\alpha\nu}B_2(Q_{\psi\gamma X)},
                \\
<\gamma 4^{++}|(yy)1>  \!&=&\!
Z^{(yy)}_{\mu\nu\lambda\sigma}p_\psi^\lambda p_\psi^\sigma
B_2(Q_{\psi\gamma X)}, \\
<\gamma 4^{++}|(yy)2> \!&=&\! g_{\mu\nu}
p_\psi^{\alpha}p_\psi^{\beta}p_\psi^{\gamma}p_\psi^{\delta}
Z^{(yy)}_{\alpha\beta\gamma\delta}B_4(Q_{\psi\gamma X)}, \\
<\gamma 4^{++}|(yy)3> \!&=&\! q_\mu
Z^{(yy)}_{\nu\lambda\sigma\alpha} p_\psi^\lambda p_\psi^\sigma
p_\psi^\alpha B_4(Q_{\psi\gamma X)}, \\
<\gamma 2^{-+}|(K^*\bar K^*)1A> \!&=&\!
\epsilon_{\mu\nu\alpha\beta}p_\psi^\alpha A^{\beta\lambda}p_{\psi\lambda}B_1(Q_{\psi\gamma X)},  \\
<\gamma 2^{-+}|(K^*\bar K^*)1B> \!&=&\!
\epsilon_{\mu\nu\alpha\beta}p_\psi^\alpha B^{\beta\lambda}p_{\psi\lambda}B_1(Q_{\psi\gamma X)},  \\
<\gamma 2^{-+}|(K^*\bar K^*)2A> \!&=&\! S_{\mu\nu}
A^{\lambda\sigma}p_{\psi\lambda}p_{\psi\sigma}B_3(Q_{\psi\gamma X)},  \\
<\gamma 2^{-+}|(K^*\bar K^*)2B> \!&=&\! S_{\mu\nu}
B^{\lambda\sigma}p_{\psi\lambda}p_{\psi\sigma}B_3(Q_{\psi\gamma X)},  \\
<\gamma 2^{-+}|(K^*\bar K^*)3A> \!&=&\! q_\mu S_{\nu\gamma}
A^{\gamma\delta}p_{\psi\delta}B_3(Q_{\psi\gamma X)},  \\
<\gamma 2^{-+}|(K^*\bar K^*)3B> \!&=&\! q_\mu\epsilon_{\nu\gamma}
B^{\gamma\delta}p_{\psi\delta}B_3(Q_{\psi\gamma X)}
\end{eqnarray}
with $S_{\mu\nu}$ defined as Eq.(\ref{smunu}). The amplitudes
involving X particles of $J^{P}=2^{+}$ involves a rank two tensor,
$X^{(yz)}$. The definition of this is given below:
\begin{eqnarray}
X_{\mu\nu}^{(\kappa\kappa)}\!&=&\! \tilde
t^{(2)}_{(\kappa\bar\kappa)\mu\nu}
f_{(14)}^{(\kappa)}f_{(23)}^{(\kappa)},\\
X_{\mu\nu}^{(K^*\bar K^*)}    \!&=&\!
P^{(2)}_{\mu\nu\alpha\beta}(K)\tilde t^{(1)\alpha}_{(14)} \tilde
t^{(1)\beta}_{(23)}  f_{(14)}^{(K^*)}f_{(23)}^{(\bar K^*)}, \\
X_{\mu\nu}^{(K^*\kappa)}    \!&=&\! P^{(2)}_{\mu\nu\alpha\beta}(K)
[\tilde t^{(1)\alpha}_{(K^*\bar\kappa)}\tilde t^{(1)\beta}_{(14)}
f_{(14)}^{(K^*)}f_{(23)}^{(\kappa)}+\tilde t^{(1)\alpha}_{(\bar
K^*\kappa)}\tilde
t^{(1)\beta}_{(23)}f_{(23)}^{(K^*)}f_{(14)}^{(\kappa)}], \\
Z^{(\kappa\kappa)}_{\alpha\beta\gamma\delta} \!&=&\! \tilde
t^{(4)}_{(\kappa\bar\kappa)\alpha\beta\gamma\delta}
f_{(14)}^{(\kappa)}f_{(23)}^{(\kappa)}, \\
Z^{(K^*\bar K^*)}_{\alpha\beta\gamma\delta} \!&=&\!
P^{(4)}_{\alpha\beta\gamma\delta\alpha'\beta'\gamma'\delta'}(K)
\tilde t^{(2)\alpha'\beta'}_{(K^*\bar K^*)}
                \tilde{t}^{(1)\gamma'}_{(14)}\tilde{t}^{(1)\delta'}_{(23)}
                f_{(14)}^{(K^*)}f_{(23)}^{(\bar K^*)}, \\
A_{\mu\nu} \!&=&\! P^{(2)}_{\mu\nu\alpha\alpha'}(K)
\epsilon^{\alpha\beta\gamma\delta} K_{\beta}
\tilde{t}^{(1)}_{(14)\gamma}\tilde{t}^{(1)}_{(23)\delta} \tilde
t^{(1)\alpha'}_{(K^*\bar K^*)}
                f_{(14)}^{(K^*)}f_{(23)}^{(\bar K^*)},\\
B_{\mu\nu} \!&=&\! P^{(2)}_{\mu\nu\alpha\alpha'}(K)
\epsilon^{\alpha\beta\gamma\delta} K_{\beta} \tilde
t^{(1)}_{(K^*\bar K^*)\gamma}
(\tilde{t}^{(1)}_{(14)\delta}\tilde{t}_{(23)}^{(1)\alpha'}
+\tilde{t}^{(1)}_{(23)\delta}\tilde{t}_{(14)}^{(1)\alpha'})
f_{(14)}^{(K^*)}f_{(23)}^{(\bar K^*)}
\end{eqnarray}
where $A_{\mu\nu}$ corresponds to $2^{-+}\to K^*\bar K^*$ with
$L=1$ and $S=1$, $B_{\mu\nu}$ corresponds to $2^{-+}\to K^*\bar
K^*$ with $L=1$ and $S=2$. We ignore $2^{-+}\to K^*\bar K^*$ with
$L=3$ due to a strong centrifugal barrier.

\section{Discussion}

Here we add some points of general technique in fitting data. The
first concerns the fact that tensor amplitudes are not always
unique. As an example, in $J/\psi \to \gamma f_2$, there are three
independent helicity amplitudes. But the general formalism allows
one to write down five covariant tensor amplitudes. Those five are
independent in the process $J/\psi \to \omega f_2$, but for the
radiative decay, gauge invariance makes two of them dependent on
the other three. Two further linear combinations differ from the
first three only by different $s$-dependence arising from the
momentum dependence built into the tensor expressions. Chung
\cite{Chung} recommends using all five combinations, so as to
retain the differences in possible $s$-dependence. However, this
gives rise to a practical problem.

\begin{figure}[htbp]
\begin{center}
\epsfysize=8.0cm \epsffile{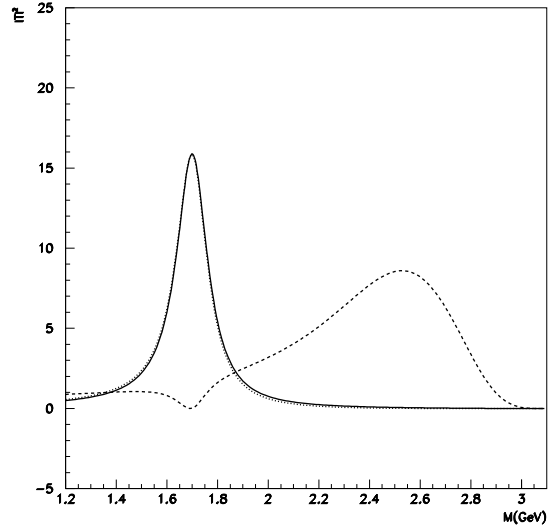}
\end{center}
\caption{\label{ptest} Distortion on Breit-Wigner amplitude
squared by the s-dependence numerators: with $Q_{\psi\gamma f}^2
B_2(Q_{\psi\gamma f})$ (solid line), with $30.3Q_{\psi\gamma f}^4
B_4(Q_{\psi\gamma f})$ (dotted line), $20[Q_{\psi\gamma f}^2
B_2(Q_{\psi\gamma f})-30.3Q_{\psi\gamma f}^4 B_4(Q_{\psi\gamma
f})$] (dashed line).}
\end{figure}

One is usually fitting resonances such as the $f_2$ to data. If
two of the amplitudes differ from the others only in
$s$-dependence, this is equivalent to putting into the numerator
of an $f_2$ Breit-Wigner amplitude a linear combination of two
s-dependent terms with two free parameters. This may lead to a
zero amplitude at the resonance mass and can give rise to
structure which may lie 500 MeV or 1 GeV away from the $f_2$; it
may be easily confused with the effects of other resonances. This
is illustrated in Fig.1 for the amplitude squared $|T|^2$ taking
as an example $J/\psi\to\gamma f_2(1700)\to\gamma K\bar K$. For
the solid line, we use $T=Q_{\psi\gamma f}^2B_2(Q_{\psi\gamma
f})/(M^2_f-s-iM_f\Gamma_f)$ with $M_f=1.7 GeV$ and $\Gamma_f=0.15
GeV$; for the dotted line which lies very close to the solid line,
we use $T=30.3*Q_{\psi\gamma f}^4B_4(Q_{\psi\gamma
f})/(M^2_f-s-iM_f\Gamma_f)$. The two different s-dependence
numerators give a hardly visible difference in line shape of
$f_2$. But if one allows two s-dependent terms in the numerator
with two free parameters, the ridiculous shape (dashed line) could
happen for a single resonance $f_2(1700)$; in this illustration we
use has $20[Q_{\psi\gamma f}^2 B_2(Q_{\psi\gamma f})-
30.3Q_{\psi\gamma f}^4 B_4(Q_{\psi\gamma f})]$ in the numerator.
Although theoretically this possibility cannot be excluded, it is
very odd and in practice one may end up fitting other $2^{++}$
components far away from the $f_2$ resonance mass with the $f_2$.
One therefore should be very careful in drawing conclusions from a
fit using more than the minimum number of amplitudes with
different angular dependence.

In $J/\psi$ radiative decays, the $c\bar c$ pair annihilates to
gluons. This requires a short-range interaction with a range of
order $1/m_c$, where $m_c$ is the mass of the $c$ quark. Therefore
the centrifugal barrier for $J/\psi\to\gamma X$ is strong. Some
production with $L=1$ is observed (at momentum transfer $\leq 1$
GeV/c), but we find little evidence for $L>1$.

\section{Acknowledgement}

We thank the Royal Society for funds allowing a collaboration
between Queen Mary College, University of London and IHEP,
Beijing. We thank V.V.Anisovich and A.V.Sarantsev for many helpful
discussions. A small part of formulae presented here were
independently written and cross checked by them. We thank
L.Y.Dong, Z.J.Guo and J.L.Hu for Monte Carlo simulation of some
formulae in this paper and for pointing out some typos. The work
is partly supported by CAS Knowledge Innovation Project
(KJCX2-SW-N02) and the National Natural Science Foundation of
China under Grant No.10225525.

\end{document}